\documentclass[a4paper,11pt]{article}
\usepackage{jcappub}
\usepackage{aas_macros}
\newcommand{\nv}{\hat{\bf n}}
\newcommand{\DMbar}{\bar{\mathcal{D}}}
\newcommand{\Dcat}{\DMbar_{\rm cat}}
\newcommand{\NF}{N_{\rm FRB}}

\newcommand{\Cov}{\mathrm{Cov}}

\title{Joint analysis of the Macquart relation and the dispersion measure power spectrum: cross-covariance and redshift self-calibration}

\author[a]{Amy Wayland}
\affiliation[a]{Department of Physics, University of Oxford, Denys Wilkinson Building, Keble Road, Oxford OX1 3RH, United Kingdom}

\author[b,c]{and Robert Reischke}
\affiliation[b]{Ruhr University Bochum, Faculty of Physics and Astronomy, Astronomical Institute (AIRUB), German Centre for Cosmological Lensing, 44780 Bochum, Germany}
\affiliation[c]{Argelander-Institut für Astronomie, Universität Bonn, Auf dem Hügel 71, D-53121 Bonn, Germany}

\emailAdd{amy.wayland@physics.ox.ac.uk}

\abstract{The dispersion measures (DMs) of fast radio bursts (FRBs) trace ionised baryons along the line of sight, probing baryonic feedback. Two complementary observables can be constructed from an FRB catalogue: the DM--redshift (Macquart) relation of host-identified FRBs, and the angular power spectrum of the DM field of the full catalogue. Their joint analysis requires their cross-covariance. We derive it within the halo model and show it is a super-sample effect: only the monopole of an individual DM correlates with the angular power spectrum, yielding an exact, model-independent bound on the correlation coefficient. We validate the calculation against log-normal forward simulations of $10^4$ FRBs. For individual DMs, the correlation with the bandpowers is at the per cent level. Neglecting the cross-covariance changes the width of the $\chi^2$ distribution by only 0.8 per cent, so the two observables can be treated as effectively independent for wide-area surveys. In a joint forecast with an LSST-like weak lensing survey, adding the Macquart relation tightens constraints on the physical baryon density by 32 per cent and on the mean host contribution by a factor of 4.1, while improving baryonification parameters by only about 5 per cent, indirectly through the host contribution. Finally, we show that the redshift distribution of FRBs without identified hosts, which enters every DM power spectrum, can be reconstructed by inverting their DMs with the jointly fitted mean relation without biasing the inferred parameters, whereas a 20 per cent error in the mean relation biases the DM power spectrum by up to 2$\sigma$ per band for $5\times 10^5$ FRBs.}

\begin{document}
\maketitle
\flushbottom

\section{Introduction} \label{sec:intro}
 Fast radio bursts (FRBs) are bright, millisecond-duration radio transients of extragalactic origin (see \cite{lorimer_bright_2007, 2022A&ARv..30....2P} 
 for a first detection and a review). As an FRB propagates towards the observer, its signal is dispersed by free electrons along the line of sight, producing a frequency-dependent arrival time delay. The dispersion measure (DM) quantifies this delay and is proportional to the column density of free electrons between the source and the observer. Since the first detections \cite{lorimer_bright_2007,Thornton2013, Petroff2015, 2016Natur.531..202S, Champion2016, Chatterjee2017}, the number of known FRBs has grown rapidly, with CHIME/FRB alone reporting hundreds of bursts in its first catalogue \cite{2021ApJS..257...59C} and a couple of thousands in its second \cite{2026ApJS..283...34C}. A growing subset of FRBs has also been localised to host galaxies with measured spectroscopic redshifts, enabling their direct use as cosmological probes. Experiments such as DSA, CHORD, and SKA \cite{Hallinan2019, Vanderlinde2020, Bacon2020} are expected to increase the total number of FRBs by two orders of magnitude, with the host-identified ones increasing from around a hundred today to many thousands.

 The sensitivity of FRB DMs to the integrated free electron density makes them a powerful probe of the distribution of ionised baryons in the Universe. Unlike observations of galaxies or neutral gas, which preferentially trace particular phases of matter, FRB DMs respond to the total column density of free electrons, including diffuse gas in filaments and in the circumgalactic and intergalactic media \cite[e.g.][]{2003ApJ...598L..79I, 2004MNRAS.348..999I,2014ApJ...780L..33M, 2014ApJ...783L..35D,2019MNRAS.485..648P}. This makes FRBs particularly sensitive to the spatially extended baryons that are difficult to detect through their electromagnetic emission alone. The distribution of these baryons is shaped by galaxy formation and feedback processes (see e.g. \cite{vanDaalen2011,somerville_physical_2015}), which can redistribute gas far beyond the regions occupied by stars. This redistribution modifies the matter power spectrum on small scales and, if not calibrated, is a leading systematic uncertainty in cosmic shear analyses \cite[e.g.][]{Chisari2019}. FRBs therefore offer a complementary means of constraining baryonic feedback, both independently and in combination with weak lensing and galaxy clustering \cite[e.g.][]{ 2026JCAP...06..006R,Reischke2025, Sharma2025, Medlock2025, Wayland2026, Sharma2026}.
 
 Two complementary statistics can be extracted from an FRB catalogue. The first is the DM--redshift relation, commonly known as the Macquart relation \cite{Macquart2020}, which is constructed from FRBs with identified host galaxies and measured redshifts. Its mean evolution with redshift probes the cosmic baryon density and the expansion history of the Universe \cite[e.g.][]{2014PhRvD..89j7303Z,hagstotz_new_2022,james_measurement_2022}. Observations of this relation have demonstrated that most of the previously ``missing'' baryons reside in the diffuse intergalactic medium \cite[e.g.][]{Macquart2020,2022ApJ...928....9L,2024ApJ...973..151K}. Beyond its mean, the scatter of the Macquart relation provides additional information about the inhomogeneous distribution of ionised gas. This scatter arises from variations in the electron column density between different sightlines and therefore depends on the distribution of baryons within and around dark matter haloes \cite[see e.g.][]{2014ApJ...780L..33M,2019MNRAS.484.1637J,2021MNRAS.505.5356B,2021ApJ...906...49Z,Medlock2025,2025NatAs...9.1226C,2025ApJ...989...81S,Reischke2025,2026OJAp....961522T}. 
 Its redshift dependence has recently been used to constrain baryonic feedback using FRBs \cite[e.g.][]{Reischke2025}.

 The second characterises the angular distribution of DMs across the sky. The angular power spectrum of the DM field, $C_\ell^{\mathcal{DD}}$, probes the projected clustering of free electrons \cite[e.g.][]{Masui2015, Shirasaki2017, Takahashi2021, Rafiei2021, Reischke2021,2026JCAP...06..006R}. Cross-correlations with other tracers of the large-scale structure can further connect the electron distribution to the underlying matter field \cite[e.g.][]{Shirasaki2017,2019PhRvD.100j3532M, Rafiei2021,Hsu2025, Hussaini2025, Sharma2025, Takahashi2026, Wayland2026}. An important advantage of angular statistics is that they do not require individual redshifts for each FRB (see e.g. \cite{2026JCAP...06..006R} for the general idea). They can therefore exploit the full catalogue, including the large majority of FRBs without an identified host, whereas the Macquart relation requires a host-identified subsample with measured redshifts. As FRB surveys continue to grow, this distinction will become increasingly important: the number of detected FRBs may grow substantially faster than the number with host associations and spectroscopic redshifts.

 Although the Macquart relation and the DM angular power spectrum probe different aspects of the electron field, their measurements are statistically related. Both are constructed from the same underlying distribution of free electrons, and the variance of the Macquart relation can be related to an integral over the DM angular power spectrum, subject to the relevant projection and selection effects. Combining the two observables can therefore exploit their complementary information, potentially improving constraints on cosmological parameters and baryonic feedback. A consistent combination must, however, account for their statistical dependence. Since both observables are measured from the same realisation of the electron field, their covariance includes a cross-covariance in addition to the covariance of each observable separately. Whereas the auto-covariances are determined by two-point statistics of the underlying field, the cross-covariance contains a connected three-point contribution sourced by the bispectrum (see \cite{2014JCAP...05..048C} for works on this position-dependent bispectrum for galaxy surveys) of the electron density field. Physically, this contribution describes the coupling between fluctuations in the line-of-sight electron column density and the angular clustering of those same electrons. Whether this coupling is sufficiently large to affect parameter inference determines whether the two probes can be analysed independently, with their information simply combined, or whether a fully joint covariance is required. The cross-covariance, arising from the squeezed bispectrum, should intuitively be very small, but to the best of our knowledge it has not previously been computed for an FRB DM catalogue.

 The combination also offers an opportunity to calibrate the redshift distribution of FRBs without individual redshift measurements. The angular power spectrum depends on the redshift distribution of the sources through its projection kernel, but most FRBs contributing to the angular statistics lack identified hosts and spectroscopic redshifts. Their redshift distribution must therefore be inferred indirectly, for example, by mapping their observed DMs to redshifts using a model of the mean Macquart relation. Errors in this relation can propagate into the inferred redshift distribution and bias the predicted DM angular power spectrum. The host-identified subsample provides independent redshift information that can constrain the Macquart relation and, in turn, calibrate the redshift distribution of the full catalogue. A joint analysis can thus improve consistency between the inferred redshift distribution and the angular statistics, while enabling us to combine the cosmological information from the host-identified population and the population without identified hosts.

 In this work, we develop a joint framework for analysing the Macquart relation and the DM angular power spectrum. We derive the full covariance of their combined data vector, calculating their cross-covariance within the halo model and under the Limber approximation. We show that the cross-covariance is analogous to super-sample covariance, allowing us to derive a model-independent bound on its amplitude and determine how it scales with survey footprint. We validate our analytical calculation against 5000 forward-simulated catalogues generated from log-normal realisations of the electron field, measuring both observables on the same realisations to retain their statistical dependence. We then assess the impact of the cross-covariance on the goodness of fit by comparing the $\chi^2$ distributions obtained using the full covariance and an approximation that neglects the cross-term, both for individual DMs and for DMs compressed into redshift bins. Finally, we use the validated covariance to forecast constraints from a joint analysis of FRB DMs and an LSST-like weak lensing survey. We quantify the improvement in parameter constraints from including the Macquart relation and investigate how a mismatch in the assumed mean relation propagates into the DM angular power spectrum through the inferred redshift distribution.

 The structure of this paper is as follows. Section~\ref{sec:theory} presents the theoretical framework, including the projection of the electron field onto the two observables, their auto-covariances, the cross-covariance, and its interpretation in terms of super-sample covariance. Section~\ref{sec:sims} describes the forward simulations and compares their covariance estimates with the analytical predictions. Section~\ref{sec:forecast} presents the forecast constraints and examines the role of the Macquart relation in calibrating the FRB redshift distribution. Finally, Section~\ref{sec:conclusions} summarises our findings and discusses their implications for future work.

\section{Theory} \label{sec:theory}
 In this section, we describe the theoretical basis for calculating the DM's statistical properties. In practice, the DM comprises different contributions: host, Milky Way and cosmological. For this discussion, we exclude these contributions because they do not (strongly) correlate with cosmological fields and would add only uncorrelated noise. Hence, in this part we work only with the cosmological contribution. Note that the literature often assigns `DM' as a variable name. We find this inhibits readability and clutters notation, so we use $\mathcal{D}$.

 \subsection{The dispersion measure as a projected field}
  The cosmological contribution to the DM of an FRB at redshift $z_1$ in direction $\nv$ is the column density of electrons along that line of sight, integrated up to that redshift. This contribution technically carries a cosmological mean (Macquart relation) and a perturbation.  
  Since the mean does not correlate with the LSS, we subtract it without loss of generality for the work here\footnote{In practice, the fact that we do not have access to the mean but have to subtract it from the \emph{observed} catalogue introduces an additional noise term and changes the survey response for spatial correlations. This is discussed in Section~\ref{ssec:theory.catalogue_pcl}.} and add it back later. In this way, the cosmological contribution may be written as a weighted projection of the electron overdensity $\delta_{\rm e}$ in direction $\nv$:
  \begin{equation} \label{eq:DM-z}
     \mathcal{D}(\nv) = \int_0^\infty \mathrm{d}\chi\, W_{\mathcal{D}}(\chi)\, \delta_{\rm e}(\chi(z)\nv, z),
  \end{equation}
  with radial kernel
  \begin{equation} \label{eq:kernel}
      W_{\mathcal{D}}(\chi) = \mathcal{A}\, (1+z(\chi)) \int_{\chi(z)}^{\chi_{\rm H}} \mathrm{d}\chi'\, n(\chi'), \qquad \mathcal{A} \equiv \frac{3 H_0^2 \Omega_{\rm b0}}{8\pi G m_{\rm p}} \, \frac{1+x_{\rm H}}{2}.
  \end{equation}
  Here, $n(\chi)$ is the comoving number density of sources, $x_{\rm H}$ is the hydrogen mass fraction, and $\chi_{\rm H}=c/H_0$. The comoving distance is denoted by $\chi$, $H_0$ is the Hubble constant, $m_\mathrm{p}$ is the proton mass, $G$ is the gravitational constant, $\Omega_\mathrm{b0}$ is the baryon density today, and $z$ is the cosmological redshift.  
  For a single FRB at redshift $z_1$, we can replace 
  \begin{equation}
      n_{z_1}(\chi') = \delta^\mathrm{D}(\chi - \chi(z_1)),
  \end{equation}
  with the Dirac delta distribution $\delta^\mathrm{D}(x)$. Alternatively, we can integrate Equation~\eqref{eq:DM-z} directly over redshift for which we write:
  \begin{equation}
      W_\mathcal{D}(z) = W_\mathcal{D}(\chi(z))\frac{\mathrm{d}z}{\mathrm{d}\chi}.
  \end{equation}
  Since the mean electron density vanishes by definition, $\langle \delta_{\rm e} \rangle = 0$, we have $\langle \mathcal{D} \rangle = 0$ for the fluctuating part, and all covariances below reduce to connected correlators of $\delta_{\rm e}$.  The corresponding angular power spectrum, assuming statistical isotropy, is
  \begin{equation} \label{eq:Cl_DD}
     C_\ell^{\mathcal{DD}}(z_1, z_2) = \frac{1}{2\ell+1} \sum_m a_{\ell m}^{\mathcal{D}}(z_1)\, \left(a_{\ell m}^{\mathcal{D}}(z_2)\right)^*,
  \end{equation}
  with the spherical harmonic coefficients
  \begin{equation} \label{eq:alm}
      a_{\ell m}^{\mathcal{D}}(z_1) = \int \mathrm{d}\hat{\mathbf{n}}\, Y_{\ell m}^*(\hat{\mathbf{n}})\, \mathcal{D}(\hat{\mathbf{n}},z_1).
  \end{equation}
  The full covariance of the joint data vector then has the block structure
  \begin{equation} \label{eq:covmat}
      \mathsf{C} = 
      \begin{pmatrix}
          \mathrm{Cov}[\mathcal{D}_i, \mathcal{D}_j] &
          \mathrm{Cov}[\mathcal{D}_i, C_\ell^{\mathcal{DD}}] \\[0.3em]
          \mathrm{Cov}[C_\ell^{\mathcal{DD}}, \mathcal{D}_j] &
          \mathrm{Cov}[C_\ell^{\mathcal{DD}}, C_{\ell'}^{\mathcal{DD}}] \\
      \end{pmatrix},
  \end{equation}
  from which we construct the correlation coefficient, which quantifies the relative importance of the off-diagonal blocks. Here we denote the DM perturbation for a single FRB as
  \begin{equation}
    \mathcal{D}_i \equiv \mathcal{D}(\nv_i, z_i),  
  \end{equation}
  with the source number density in Equation~\eqref{eq:kernel} replaced by a delta distribution at the FRB's comoving distance. Note again that the observable in the upper block is the Macquart relation, so the perturbation and the mean. For simplicity, we will keep the notation general with $W_\mathcal{D}$, with the above identification in mind. The perturbations, however, fully capture the covariance structure. The following subsections derive the three blocks in turn. 

 \subsection{Auto-covariance of the DM--redshift relation}
  The auto-covariance between two FRBs $i$ and $j$ is 
  \begin{equation}
     \mathrm{Cov}[\mathcal{D}(\hat{\mathbf{n}}_i, z_i), \mathcal{D}(\hat{\mathbf{n}}_j, z_j)] = \left\langle \mathcal{D}(\hat{\mathbf{n}}_i, z_i) \mathcal{D}(\hat{\mathbf{n}}_j, z_j) \right\rangle - \left\langle \mathcal{D}(\hat{\mathbf{n}}_i, z_i) \right\rangle \left\langle \mathcal{D}(\hat{\mathbf{n}}_j, z_j) \right\rangle,
  \end{equation}
  where the second term vanishes as noted above. Substituting Equation~\eqref{eq:DM-z} yields
  \begin{equation}
     \mathrm{Cov}[\mathcal{D}_i, \mathcal{D}_j] = \int_0^{z_i} \mathrm{d}z_i' \, W_{\mathcal{D}}(z_i') \int_0^{z_j} \mathrm{d}z_j' \, W_{\mathcal{D}}(z_j') \, \left\langle \delta_{\rm e}(\chi_i'\hat{\mathbf{n}}_i, z_i') \, \delta_{\rm e}(\chi_j'\hat{\mathbf{n}}_j, z_j') \right\rangle,
  \end{equation}
  with the electron power spectrum defined by
  \begin{equation}
     \left\langle \delta_{\rm e}(\mathbf{k},z_i) \, \delta_{\rm e}^*(\mathbf{k}', z_j) \right\rangle = (2\pi)^3 \delta^{\rm D}(\mathbf{k}-\mathbf{k}') P_{\rm e}(k; z_i, z_j).
  \end{equation}
  Expanding the plane waves in spherical harmonics (see Equation~\ref{eq:rayleigh_expansion}), the auto-covariance between the FRBs becomes \cite{Reischke2023}
  \begin{equation} \label{eq:covDD_legendre}
     \mathrm{Cov}[\mathcal{D}_i, \mathcal{D}_j] = \sum_\ell \frac{2\ell+1}{4\pi} \, P_\ell(\hat{\mathbf{n}}_i \cdot \hat{\mathbf{n}}_j) \, C_{ij}(\ell),
  \end{equation}
  where $\hat{\mathbf{n}}_i \cdot \hat{\mathbf{n}}_j = \cos\theta_{ij}$ is the cosine of the angular separation between the two sightlines and
  \begin{align}
  \label{eq:Cij}
      C_{ij}(\ell) = \frac{2}{\pi} \int_0^\infty \mathrm{d}k \, k^2 \left[\int_0^{z_i} \mathrm{d}z_i' \, W_\mathcal{D}(z_i') \sqrt{P_{\rm e}(k, z_i')} j_\ell(k\chi_i')\right]& \nonumber \\[0.2em]
      \times\left[\int_0^{z_j} \mathrm{d}z_j' \, W_\mathcal{D}(z_j') \sqrt{P_{\rm e}(k, z_j')} j_\ell(k\chi_j')\right]&,
  \end{align}
  where $j_\ell(x)$ is a spherical Bessel function of order $\ell$.
  Here, we have used the approximation $P_{\rm e}(k; z_i, z_j) \approx \sqrt{P_{\rm e}(k, z_i) P_{\rm e}(k, z_j)}$, which is valid for broad projection kernels~\cite{Kitching2017}. Applying the Limber approximation \cite{2008PhRvD..78l3506L},
  \begin{equation} \label{eq:limber}
      j_{\ell}(k\chi) \approx \sqrt{\frac{\pi}{2\ell+1}} \, \delta^{\rm D}\left(k - \frac{\ell+1/2}{\chi}\right) \frac{1}{{\chi}},
  \end{equation}
  collapses the integral over $k$, giving
  \begin{equation} \label{eq:Cij_limber}
      C_{ij}(\ell) = \int_0^{\rm min(\chi_i, \chi_j)} \frac{\mathrm{d}\chi}{\chi^2} \, W_\mathcal{D}^2(\chi) \, P_{\rm e}\left(\frac{\ell+1/2}{\chi}, z(\chi)\right).
  \end{equation}
  The upper integration limit reflects the fact that only the overlapping portion of the two lines of sight contributes to the correlation. Note that quoting $ \mathrm{Cov}[\mathcal{D}_i, \mathcal{D}_j]$ is \emph{not} a statement about the corresponding likelihood.
 
 \subsection{Auto-covariance of the DM angular power spectrum}
  The covariance of the DM angular power spectrum can, by virtue of Wick's theorem and ignoring mode-coupling due to masking, be written as the sum of a Gaussian and a non-Gaussian part,
  \begin{equation} \label{eq:cov_total}
      \mathrm{Cov}\left[C_{\ell_1}^{\mathcal{DD}}, C_{\ell_2}^{\mathcal{DD}}\right] = \mathrm{Cov}^{\rm G} + \mathrm{Cov}^{\rm cNG}.
  \end{equation}
  The Gaussian part is described by the Knox formula \cite{Knox1997}, given by
  \begin{equation} \label{eq:knox}
      \mathrm{Cov}[{C}_\ell^{\mathcal{DD}}, C_{\ell'}^{\mathcal{DD}}] = \frac{2\delta_{\ell\ell'}^{\rm K}}{(2\ell+1) f_{\rm sky}} \, (\hat{C}_\ell^{\mathcal{DD}})^2,
  \end{equation}
  where $f_{\rm sky}$ is the observed sky fraction and $\hat{C}^{\mathcal{DD}}_\ell$ the observed angular power spectrum:
  \begin{equation}
      \hat{C}_\ell^{\mathcal{DD}} = {C}_\ell^{\mathcal{DD}} + \frac{\sigma^2_\mathcal{D}}{\bar n}\;,
  \end{equation}
  with the number density of FRBs $\bar n$ and the field variance $\sigma^2_\mathcal{D}$, i.e. host variance plus the variance induced by the large-scale structure.
  The Gaussian covariance is therefore diagonal in $\ell$, reflecting the statistical independence of multipoles. The full-sky DM angular power spectrum entering Equation~\eqref{eq:knox} is
  \begin{equation}
      C_\ell^{\mathcal{DD}} = \int_0^{\chi_{\rm H}} \frac{\mathrm{d}\chi}{\chi^2} \, W_{\mathcal{D}}^2(\chi) \, P_{\rm e}\left(\frac{\ell+1/2}{\chi}, z(\chi)\right).
  \end{equation}
  By contrast, the non-Gaussian part arises from the connected part of the four-point function, sourced by the trispectrum of the electron density field. We neglect super-sample covariance from partial sky areas \cite{2013PhRvD..87l3504T,2014PhRvD..89h3519L}, as it is reduced for the DM because we subtract the mean contribution, as in galaxy clustering \cite{2017MNRAS.470.2100K}.
  Under the Limber approximation, this reads
  \begin{equation} \label{eq:cng_limber}
      \mathrm{Cov}^{\rm cNG}(\ell_1,\ell_2) = \frac{1}{4\pi f_{\rm sky}} \int\frac{\mathrm{d}\chi}{\chi^6}\, W_{\mathcal{D}}^4(\chi)\, \bar{T}_{\rm eeee}\left[\frac{\ell_1+1/2}{\chi}, \frac{\ell_2+1/2}{\chi}, a(\chi)\right],
  \end{equation}
  where $\bar{T}_{\rm eeee}$ is the isotropised connected trispectrum of $\delta_{\rm e}$.

  The angular power spectrum is measured in bands rather than at individual multipoles, so we must convert both terms of Equation~\eqref{eq:cov_total} into a bandpower covariance. Following Ref.~\cite{Reischke2023}, a bandpower is defined with the measure $\ell\,\mathrm{d}\ell$,
  \begin{equation} \label{eq:bandpower}
      \bar{C}_b = \frac{\int_b \mathrm{d}\ell\, \ell\, C_\ell}{\int_b \mathrm{d}\ell\, \ell} \to \frac{\sum_{\ell \in b} \ell\, C_\ell}{\sum_{\ell \in b} \ell},
  \end{equation}
  where the arrow gives the discrete form used for integer multipoles on the full sky, and the band $b$ is a top hat $[\ell_{\rm lo},\ell_{\rm hi})$ so that the band window is absorbed into the summation limits. Because Equation~\eqref{eq:bandpower} is linear in $C_\ell$, the covariance of two bandpowers follows directly
  \begin{equation} \label{eq:band_cov}
      \mathrm{Cov}\left[\bar{C}_{b_1}, \bar{C}_{b_2}\right] = \frac{\sum_{\ell_1 \in b_1} \sum_{\ell_2 \in b_2} \ell_1 \ell_2\, \mathrm{Cov}[C_{\ell_1}, C_{\ell_2}]}{\sum_{\ell_1 \in b_1} \ell_1 \; \sum_{\ell_2 \in b_2} \ell_2},
  \end{equation}
  which is Equation~(47) of Ref.~\cite{Reischke2023} in our notation. The weights in Equation~\eqref{eq:band_cov} are the same for the Gaussian and the connected non-Gaussian term. 
  
  The two terms behave differently under Equation~\eqref{eq:band_cov}. The Gaussian term is diagonal in multipole, so the Kronecker delta collapses the double sum,
  \begin{equation} \label{eq:covb1b2_g}
      {\rm Cov}^{\rm G}[\bar{C}_b,\bar{C}_{b'}] = \delta_{bb'}^{\mathrm{K}} \frac{\sum_{\ell \in b} \frac{2 C_\ell^2}{(2\ell+1)f_{\rm sky}}}{(\sum_{\ell \in b}\ell)^2} \simeq \delta_{bb'}^{\mathrm{K}} \frac{2 \bar{C}_b^2}{f_{\rm sky}\, N_b}, \qquad N_b \equiv \sum_{\ell \in b}(2\ell+1),
  \end{equation}
  where the last step assumes that $C_\ell$ varies slowly across the band. The connected term, by contrast, is a smooth and positive function of both arguments and is strongly correlated between neighbouring multipoles, so Equation~\eqref{eq:band_cov} simply returns a weighted mean of ${\rm Cov}^{\rm cNG}(\ell,\ell')$ over the band pair,
  \begin{equation} \label{eq:covb1b2_cng}
      \mathrm{Cov}^{\rm cNG}_{b_1 b_2} \simeq \mathrm{Cov}^{\rm cNG}(\ell_{b_1},\ell_{b_2}),
  \end{equation}
  where $\ell_b$ is a representative multipole of band $b$. Combining Equations~\eqref{eq:covb1b2_g} and~\eqref{eq:covb1b2_cng}, the ratio of the two contributions in a band is related to the ratio at a single multipole by
  \begin{equation} \label{eq:cng_to_g_ratio}
      \frac{{\rm Cov}^{\rm cNG}}{{\rm Cov}^{\rm G}} \simeq \left.\frac{{\rm Cov}^{\rm cNG}(\ell,\ell)}{{\rm Cov}^{\rm G}(\ell,\ell)} \right|_{\ell=\ell_b} \times \frac{N_b}{2\ell_b+1} \sim \left.\frac{{\rm Cov}^{\rm cNG}(\ell,\ell)}{{\rm Cov}^{\rm G}(\ell,\ell)} \right|_{\ell=\ell_b} \times \Delta\ell_b,
  \end{equation}
  where $\Delta\ell_b = \ell_{\rm hi}-\ell_{\rm lo}$ is the bandwidth. For the logarithmic bands used here, $\Delta\ell_b/\ell_b \simeq 0.5$, so the band averaging enhances the relative importance of the connected term by a factor that grows from $\sim1$ at $\ell\simeq5$ to several hundred at $\ell\simeq10^3$. A connected term that is subdominant at every individual multipole can therefore dominate the bandpower covariance at high $\ell$. In practice, however, the shot noise will dominate.

 \subsection{Cross-covariance of the two observables} \label{sec:cross-cov}
  We now derive the cross-covariance between the DM--redshift relation and the DM angular power spectrum,
  Using $\mathrm{Cov}[X,YZ] = \langle XYZ \rangle - \langle X \rangle \langle YZ \rangle$, this may be written as
  \begin{equation}
      \mathrm{Cov}\left[\mathcal{D}(\hat{\mathbf{n}}_1, z_1), \,C_\ell^{\mathcal{DD}}(z_2, z_3)\right] = \left\langle \mathcal{D}(\hat{\mathbf{n}}_1, z_1) \, C_\ell^{\mathcal{DD}}(z_2, z_3) \right\rangle - \left\langle \mathcal{D}(\hat{\mathbf{n}}_1, z_1) \right\rangle  \left\langle C_\ell^{\mathcal{DD}}(z_2, z_3) \right\rangle.
  \end{equation}
  Since $\left\langle \mathcal{D}(\hat{\mathbf{n}}, z_1) \right\rangle = 0$, the second contribution vanishes. The expression therefore reduces to the connected three-point function of the DM field:
  \begin{align}
      \mathrm{Cov}\left[\mathcal{D}(\hat{\mathbf{n}}_1, z_1), \,C_\ell^{\mathcal{DD}}(z_2, z_3)\right] &= \left\langle \mathcal{D}(\hat{\mathbf{n}}_1, z_1) \, C_\ell^{\mathcal{DD}}(z_2, z_3) \right\rangle \nonumber \\[0.5em]
      &= \frac{1}{2\ell+1} \sum_m \left\langle \mathcal{D}(\hat{\mathbf{n}}_1, z_1) \, a_{\ell m}^{\mathcal{D}}(z_2) \, \left(a_{\ell m}^{\mathcal{D}}(z_3)\right)^* \right\rangle .
  \end{align}
  To evaluate it, we follow steps similar to Ref.~\cite{Reischke2023}. Substituting Equation~\eqref{eq:DM-z} into Equation~\eqref{eq:alm} gives
  \begin{equation}
      a_{\ell m}^{\mathcal{D}}(z_1) = \int_0^{z_1} \mathrm{d}z \, W_\mathcal{D}(z) \int \mathrm{d}\hat{\mathbf{n}} \, Y_{\ell m}^*(\hat{\mathbf{n}}) \, \delta_{\rm e}(\chi(z)\hat{\mathbf{n}},z).
  \end{equation}
  Writing the electron overdensity as a Fourier integral lets us express the spherical harmonic coefficients of $\mathcal{D}$ (see Appendix~\ref{app:spherical_harmonic_basics}). 
    Using the definition of the bispectrum
  \begin{equation}
      \left\langle \delta_{\rm e}(\mathbf{k}_1,z_1) \, \delta_{\rm e}(\mathbf{k}_2,z_2) \, \delta_{\rm e}^*(\mathbf{k}_3,z_3) \right\rangle = (2\pi)^3 \delta^\mathrm{D}(\mathbf{k}_1+\mathbf{k}_2-\mathbf{k}_3) \, B_{\rm e}(k_1, k_2, k_3 ; z_1, z_2, z_3),
  \end{equation}
  allows us to write the covariance as
  \begin{align}
      \mathrm{Cov}\left[\mathcal{D}(\hat{\mathbf{n}}_1, z_1), \,C_\ell^{\mathcal{DD}}\right] &= \frac{(4\pi)^2}{2\ell+1} \sum_m \int_0^{z_1} \mathrm{d}z_1' \, W_{\mathcal{D}}(z_1') \int_0^{\infty} \mathrm{d}z_2' \, W_{\mathcal{D}}(z_2') \int_0^{\infty} \mathrm{d}z_3' \, W_{\mathcal{D}}(z_3') \nonumber \\
      &\times \int \frac{\mathrm{d}^3 k_1}{(2\pi)^3} \int \frac{\mathrm{d}^3 k_2}{(2\pi)^3} \, e^{\mathrm{i} \mathbf{k}_1 \cdot (\chi(z_1') \hat{\mathbf{n}}_1)} \, j_{\ell}(k_2 \chi(z_2')) \, j_{\ell}(k_3 \chi(z_3')) \nonumber \\[0.5em]
      &\times Y_{\ell m}^*(\hat{\mathbf{k}}_2) \, Y_{\ell m}(\hat{\mathbf{k}}_3)  \, B_{\rm e}(k_1, k_2, \mu_{12} ; z_1', z_2', z_3'),
  \end{align}
  where $\mathbf{k}_3 = \mathbf{k}_1 + \mathbf{k}_2$ and $\mu_{ij} = \cos\theta_{ij} = \hat{\mathbf{k}}_i \cdot \hat{\mathbf{k}}_j$. Using Equation \eqref{app:eq:legendre} collapses the sum over $m$ and simplifies the prefactor, giving
  \begin{align}
      \mathrm{Cov}\left[\mathcal{D}(\hat{\mathbf{n}}_1, z_1), \,C_\ell^{\mathcal{DD}}\right] &= 4\pi \int_0^{z_1} \mathrm{d}z_1' \, W_{\mathcal{D}}(z_1') \int_0^{\infty} \mathrm{d}z_2' \, W_{\mathcal{D}}(z_2') \int_0^{\infty} \mathrm{d}z_3' \, W_{\mathcal{D}}(z_3') \nonumber \\
      &\times \int \frac{\mathrm{d}^3 k_1}{(2\pi)^3} \int \frac{\mathrm{d}^3 k_2}{(2\pi)^3} \, e^{\mathrm{i} \mathbf{k}_1 \cdot (\chi(z_1') \hat{\mathbf{n}}_1)} \, j_{\ell}(k_2 \chi(z_2')) \, j_{\ell}(k_3 \chi(z_3')) \nonumber \\[0.5em]
      &\times P_\ell(\hat{\mathbf{k}}_2 \cdot \hat{\mathbf{k}}_3)  \, B_{\rm e}(k_1, k_2, \mu_{12} ; z_1', z_2', z_3').
  \end{align}
  The cross-covariance between the Macquart relation and the DM angular power spectrum therefore arises from the bispectrum $B_{\rm e}(k_1, k_2, \mu_{12})$ of the electron overdensity field. Using the Rayleigh expansion, Equation \eqref{eq:rayleigh_expansion}, yields
  \begin{align}
      \mathrm{Cov}\left[\mathcal{D}(\hat{\mathbf{n}}_1, z_1), \,C_\ell^{\mathcal{DD}}\right] &= (4\pi)^2 \int_0^{z_1} \mathrm{d}z_1' \, W_{\mathcal{D}}(z_1') \int_0^{\infty} \mathrm{d}z_2' \, W_{\mathcal{D}}(z_2') \int_0^{\infty} \mathrm{d}z_3' \, W_{\mathcal{D}}(z_3') \nonumber \\
      &\times \sum_{LM} \mathrm{i}^L Y_{LM}(\hat{\mathbf{n}}_1) \nonumber \\
      &\times \int \frac{\mathrm{d}^3 k_1}{(2\pi)^3} \int \frac{\mathrm{d}^3 k_2}{(2\pi)^3} \, j_L(k_1 \chi(z_1')) \, j_{\ell}(k_2 \chi(z_2')) \, j_{\ell}(k_3 \chi(z_3')) \nonumber \\[0.5em]
      &\times Y_{LM}^*(\hat{\mathbf{k}}_1) \, P_\ell(\hat{\mathbf{k}}_2 \cdot \hat{\mathbf{k}}_3) \,  B_{\rm e}(k_1, k_2, \mu_{12} ; z_1', z_2', z_3').
  \end{align}
  On the full sky, isotropy requires that the result not depend on the direction $\hat{\mathbf{n}}_1$ in which $\mathcal{D}$ is measured, so we can average over it. Using
  \begin{equation} \label{eq:Ylm_average}
      \frac{\int\mathrm{d}\Omega \, Y_{\ell m}(\hat{\mathbf{n}}_1)}{4\pi} = \frac{1}{\sqrt{4\pi}}\delta^\mathrm{K}_{\ell0}\,\delta^\mathrm{K}_{m0},
  \end{equation}
  to collapse the sum:
  \begin{align}
  \label{eq:bessel}
      \mathrm{Cov}\left[\mathcal{D}(z_1), \,C_\ell^{\mathcal{DD}}\right] &= \frac{4\pi}{(2\pi)^6} \int_0^{z_1} \mathrm{d}z_1' \, W_{\mathcal{D}}(z_1') \int_0^{\infty} \mathrm{d}z_2' \, W_{\mathcal{D}}(z_2') \int_0^{\infty} \mathrm{d}z_3' \, W_{\mathcal{D}}(z_3') \nonumber \\
      &\times \int \mathrm{d}^3 k_2 \int {\mathrm{d}^3 k_3} \, j_0(|-\mathbf{k}_2 + \mathbf{k}_3| \chi(z_1')) \, j_{\ell}(k_2 \chi(z_2')) \, j_{\ell}(k_3 \chi(z_3')) \nonumber \\[0.5em]
      &\times  \, P_\ell(\hat{\mathbf{k}}_2 \cdot \hat{\mathbf{k}}_3)  \, B_{\rm e}(k_2, k_3, \mu_{23} ; z_1', z_2', z_3').
  \end{align}
  This is a crucial step, and we revisit it in Section \ref{sec:cross_partial} to investigate the effects of partial sky coverage.
  Next, we rotate the coordinate system so that it points along $\mathbf{k}_2$, allowing us to integrate over $\mathbf{k}_2$ and obtain a factor of $4\pi$. Similarly, we perform the polar angle integral over $\mathbf{k}_3$ and apply the Limber approximation, given by Equation~\eqref{eq:limber}, to obtain the final result:
  \begin{align} \label{eq:crosscov_final}
      \mathrm{Cov}\left[\mathcal{D}(z_1), \,C_\ell^{\mathcal{DD}}\right] &= \frac{(\ell +\tfrac{1}{2})^3}{4\pi^2\chi_{H}^3} \int_0^{\chi(z_1)} {\mathrm{d}\chi_1'}{E(\chi_1')} \, W_{\mathcal{D}}(z(\chi_1')) \int_0^{\chi_H} \frac{\mathrm{d}\chi_2'}{\chi'^3_2} \,E(\chi_2')\, W_{\mathcal{D}}(z(\chi_2'))  \nonumber \\ &\times \int_0^{\chi_H} \frac{\mathrm{d}\chi_3'}{\chi'^3_3}E(\chi_3')\, W_{\mathcal{D}}(z(\chi_3'))
       \int_{-1}^1\mathrm{d}\mu_{12}\, j_0\left(f\left[\frac{\ell +\tfrac{1}{2}}{\chi_2'}, \frac{\ell +\tfrac{1}{2}}{\chi_3'},\mu_{12}\right] \chi_1'\right)\;  \nonumber \\ &\times   P_\ell(\mu_{12})  \, B_{\rm e}\left(\frac{\ell +\tfrac{1}{2}}{\chi_2'}, \frac{\ell +\tfrac{1}{2}}{\chi_3'}, \mu_{12} ; \chi_1', \chi_2', \chi_3'\right),
  \end{align}
 where $f^2[k_1,k_2,\mu_{12}] = k_1^2 +k_2^2-2\mu_{12} k_1 k_2 $. For the specific computations of the halo-model ingredients, we refer to Appendix~\ref{app:sec:halo_model}. For the unequal-time correlator, we assume a geometric mean as in Equation~\eqref{eq:Cij}.
 
 \subsection{Super-sample origin and partial sky coverage} \label{ssec:theory.supersample}
 \label{sec:cross_partial}
  The result of Section~\ref{sec:cross-cov} assumes a statistically isotropic field observed over the full sky. Here, we make the origin of the cross-covariance explicit, show that it is a super-sample effect, and generalise it to a survey window. 

  \subsubsection{The full sky bound}
   The average over $\nv_1$ in Equation~\eqref{eq:Ylm_average} has a clear physical interpretation. Since $C_\ell^{\mathcal{DD}}$ is invariant under rotations and the field is statistically isotropic, $\mathrm{Cov}[\mathcal{D}(\nv, z_i), C_\ell^{\mathcal{DD}}]$ does not depend on $\nv$ and is therefore equal to its sky average,
   \begin{equation} \label{eq:cross_monopole}
       \mathrm{Cov}\left[\mathcal{D}(\nv_i, z_i), \,C_\ell^{\mathcal{DD}}\right] = \mathrm{Cov}\left[\mathcal{D}_\Omega(z_i), \,C_\ell^{\mathcal{DD}}\right], \qquad
       \mathcal{D}_\Omega(z_i) \equiv \int\frac{\mathrm{d}\Omega}{4\pi}\,\mathcal{D}(\nv, z_i) = \frac{a_{00}^{\mathcal{D}}(z_i)}{\sqrt{4\pi}}.
   \end{equation}
   Although a single FRB's DM contains all multipoles, only its monopole projection correlates with the angular power spectrum. Writing $\mathcal{D}_i = \mathcal{D}_\Omega(z_i) + \mathcal{D}_{\perp,i}$, Equation~\eqref{eq:cross_monopole} implies $\mathrm{Cov}[\mathcal{D}_{\perp,i}, C_\ell^{\mathcal{DD}}] = 0$. The Cauchy--Schwarz inequality, together with Equation~\eqref{eq:covDD_legendre} evaluated at zero separation, then bounds the correlation coefficient between $\mathcal{D}_i$ and $C_\ell^{\mathcal{DD}}$:
   \begin{equation} \label{eq:r_bound_fullsky}
       |r| \leq \frac{\sigma\left(\mathcal{D}_\Omega(z_i)\right)}{\sigma(\mathcal{D}_i)} = \left[\frac{C_{ii}(0)}{\sum_\ell (2\ell+1)\, C_{ii}(\ell)}\right]^{1/2}.
   \end{equation}
   This bound is exact in the field, regardless of the bispectrum amplitude. Because the variance of a single DM is dominated by high multipoles, the monopole fraction is small, and the cross-covariance is consequently suppressed.

  \subsubsection{Partial sky and super-sample correspondence}
   In Equation~\eqref{eq:bessel}, the argument of $j_0$ is the wavenumber $k_1 = f[k_2,k_3,\mu_{12}]$, so the integral is dominated by $k_1 \lesssim 1/\chi_1'$, while $k_2 \simeq k_3 \simeq (\ell+1/2)/\chi$. The bispectrum is therefore probed in the squeezed limit, where it is described by the response of the small-scale electron power spectrum to a long-wavelength matter mode $\delta_{\rm b}$ (often referred to as a background mode in the spirit of a separate Universe approach \cite{2015MNRAS.448L..11W}),
   \begin{equation} \label{eq:response_def}
       P_{\rm e}(k, z\,|\,\delta_{\rm b}) = P_{\rm e}(k,z)\left[1 + R_{\rm e}(k,z)\,\delta_{\rm b}\right], \qquad R_{\rm e} \equiv \frac{\partial \ln P_{\rm e}}{\partial \delta_{\rm b}}.
   \end{equation}
   In the halo model, this can be calculated at tree level:
   \begin{equation} \label{eq:response_halomodel}
       \frac{\partial P_{\rm e}}{\partial \delta_{\rm b}}(k,z) = \left[\frac{68}{21} - \frac{1}{3}\frac{\mathrm{d}\ln\left(k^3 P_{\rm lin}(k,z)\right)}{\mathrm{d}\ln k}\right]\left[I_1^1(k,z)\right]^2 P_{\rm lin}(k,z) + I_2^1(k,k,z).
   \end{equation}
   This leads to a perturbation to the angular power spectrum in the Limber approximation containing:
   \begin{equation} \label{eq:Cl_response}
    \begin{split}
       \delta C_\ell^{\mathcal{DD}}&  \supset \int\mathrm{d}\chi\, K_\ell(\chi)\, R_{\rm e}\left(\frac{\ell+1/2}{\chi}, z(\chi)\right)\delta_{\rm b}(\chi), \\
       K_\ell(\chi)&  \equiv \frac{W_{\mathcal{D}}^2(\chi)}{\chi^2}\, P_{\rm e}\left(\frac{\ell+1/2}{\chi}, z(\chi)\right).
    \end{split}
   \end{equation}
   On the full sky,
   \begin{equation}
       \delta_{\rm b}(\chi) = \int\mathrm{d}\Omega\,\delta(\chi\nv)/4\pi
   \end{equation}
   is the monopole of the matter overdensity on the shell at distance $\chi$. On these scales, the electron overdensity traces the matter with unit bias, as the linear growth equation is the same for dark matter and electrons on large scales (in other words: mass conservation). We define the shell spectra of those long modes as
   \begin{equation} \label{eq:shell_CL}
       \mathcal{C}_L(\chi, \chi') = \frac{2}{\pi}\int_0^\infty \mathrm{d}k\,k^2\,P_{\rm lin}(k; z, z')\, j_L(k\chi)\, j_L(k\chi'),
   \end{equation}
   where $P_{\rm lin}$ is the linear matter power spectrum, and denoting $z = z(\chi)$ and $z' = z(\chi')$. Applying the same unequal-time approximation as in Equation~\eqref{eq:Cij} yields
   \begin{equation} \label{eq:cross_response_fullsky}
       \mathrm{Cov}\left[\mathcal{D}(z_i), \,C_\ell^{\mathcal{DD}}\right] \simeq \frac{1}{4\pi}\int_0^{z_i}\mathrm{d}z'\, W_{\mathcal{D}}(z')\, \int\mathrm{d}\chi\, K_\ell(\chi)\, R_{\rm e}\left(\frac{\ell+1/2}{\chi}, z(\chi)\right)\mathcal{C}_0(\chi(z'), \chi).
   \end{equation}
   Thus, the cross-covariance is the analogue of super-sample covariance \cite{2006MNRAS.371.1188H,2013PhRvD..87l3504T}: the DM of an FRB measures the background mode, and the power spectrum responds to it. Equation~\eqref{eq:crosscov_final} evaluates the full bispectrum, with Equation~\eqref{eq:cross_response_fullsky} as its squeezed limit \cite{2014JCAP...05..048C,2015JCAP...08..042W}. For a survey window $w(\nv)$ with 
   \begin{equation}
       f_{w^2} = \frac{1}{4\pi}\int\mathrm{d}\Omega\,w^2(\nv),
   \end{equation} 
   the long mode seen by the pseudo-$C_\ell$ estimator, $\tilde{C}_\ell$, is the window-weighted mean,
   \begin{equation} \label{eq:deltab_window}
       \delta_\mathrm{b}^w(\chi) = \frac{1}{4\pi f_{w^2}}\int\mathrm{d}\Omega\, w^2(\nv)\,\delta(\chi\nv),
   \end{equation}
   with the linear density contrast $\delta$. Expanding the correlation function in Legendre polynomials and using the addition theorem,
  \begin{equation} \label{eq:Xi_window}
       \left\langle \delta_{\rm e}(\chi'\nv, z')\,\delta_\mathrm{b}^w(\chi)\right\rangle = \frac{1}{4\pi f_{w^2}}\sum_{LM}\mathcal{C}_L(\chi',\chi)\,\left(w^2\right)_{LM}\,Y_{LM}(\nv),
  \end{equation}
   where $(w^2)_{LM}$ are the spherical harmonic coefficients of $w^2$. Alternatively, in a more convenient form, it can be written in real space as:
   \begin{equation} \label{eq:Xi_real}
       \Xi(\nv;\chi',\chi) = \frac{1}{4\pi f_{w^2}}\int\mathrm{d}\Omega'\, w^2(\nv')\,\sum_L \frac{2L+1}{4\pi}\,\mathcal{C}_L(\chi',\chi)\,P_L(\nv\cdot\nv').
   \end{equation}
   The cross-covariance is therefore given by
   \begin{equation} \label{eq:cross_response_window}
       \mathrm{Cov}\left[\mathcal{D}(\nv_i, z_i), \,\tilde{C}_\ell\right] \simeq \int_0^{z_i}\mathrm{d}z'\, W_{\mathcal{D}}(z') \int\mathrm{d}\chi\, K_\ell(\chi)\, R_{\rm e}\left(\frac{\ell+1/2}{\chi}, z(\chi)\right)\Xi(\nv_i,\chi(z'), \chi),
   \end{equation}
  which reduces to Equation~\eqref{eq:cross_response_fullsky} for $w=1$. This result can be compared with the super-sample covariance contribution to the covariance of the angular power spectrum \cite{2018JCAP...06..015B,2019A&A...624A..61L},
   \begin{equation} \label{eq:SSC_window}
       \mathrm{Cov}^{\rm SSC}\left[\tilde{C}_{\ell}, \tilde{C}_{\ell'}\right] \simeq \iint\mathrm{d}\chi\,\mathrm{d}\chi'\, K_\ell R_{\rm e}\, K_{\ell'} R_{\rm e}\,\frac{1}{(4\pi f_{w^2})^2}\sum_L \mathcal{C}_L(\chi,\chi')\sum_M\left|\left(w^2\right)_{LM}\right|^2.
   \end{equation}
   The cross-covariance in Equation~\eqref{eq:cross_response_window} has the same structure as the super-sample covariance contribution, with one of the two survey windows replaced by the pencil beam of the FRB, $(w^2)^*_{LM}/(4\pi f_{w^2}) \to Y^*_{LM}(\nv_i)$, and the corresponding response replaced by the DM kernel.

  \subsubsection{Bounding the partial sky expression}
   The bound of Equation~\eqref{eq:r_bound_fullsky} relies on statistical isotropy. Within the response approximation, only the contribution in Equation~\eqref{eq:Cl_response} correlates with $\mathcal{D}_i$. To generalise this bound to partial sky coverage, we define the window-weighted mean DM as
   \begin{equation}
       \mathcal{D}_w(z_i) = \frac{1}{4\pi f_{w^2}}\int\mathrm{d}\Omega\,w^2\,\mathcal{D}(\nv, z_i).
   \end{equation} 
   For an FRB located well within a survey footprint of area $\Omega_{\rm s} = 4\pi f_{\rm sky}$, the covariance with this window-weighted mean is approximately $\mathrm{Cov}[\mathcal{D}_i, \mathcal{D}_w] \simeq \sigma_\mathrm{b}^2$, where
   \begin{equation} \label{eq:sigmab}
       \sigma_\mathrm{b}^2 = \frac{1}{(4\pi f_{w^2})^2}\sum_L C_{ii}(L)\sum_M\left|\left(w^2\right)_{LM}\right|^2 \simeq \frac{C_{ii}(L\lesssim L_{\rm s})}{\Omega_{\rm s}},
   \end{equation}
   and $ L_{\rm s} \sim {\pi}/{\sqrt{\Omega_{\rm s}}}$ characterises the angular scale of the survey footprint. In the last step, we have used Parseval's theorem, assumed a compact top-hat window, and taken $C_{ii}(L)$ to be approximately constant for $L \lesssim L_{\rm s}$. The resulting bound on the cross-correlation coefficient is then
   \begin{equation} \label{eq:r_bound_window}
       |r| \lesssim \frac{\sigma_\mathrm{b}}{\sigma(\mathcal{D}_i)} \simeq \left[\frac{C_{ii}(L\lesssim L_{\rm s})}{f_{\rm sky}\sum_L(2L+1)\,C_{ii}(L)}\right]^{1/2} \propto f_{\rm sky}^{-1/2}.
   \end{equation}
   Thus, the cross-correlation is enhanced as the footprint shrinks, following the same qualitative scaling as the super-sample covariance contribution relative to the Gaussian covariance. For FRBs outside the footprint, the cross-correlation instead decreases with angular separation on the characteristic correlation scale of the long-wavelength mode. Note that this bound assumes that the mean of the field is subtracted on an object-by-object basis. This does not technically hold, as we will discuss in the next section.

 \subsection[Covariance of the catalogue pseudo-$C_\ell$ estimator]{\boldmath The catalogue pseudo-$C_\ell$ estimator} \label{ssec:theory.catalogue_pcl}
  An FRB catalogue provides the DM measurements along $\NF$ discrete sightlines. We denote the index set of all FRBs by $\mathcal{S}$ and the subset of the $N_{\rm host}$ FRBs with identified hosts by $\mathcal{L}\subset \mathcal{S}$. Following Refs.~\cite{Alonso2019,2024JCAP...05..010B,2025JCAP...01..028W}, we estimate the angular power spectrum directly from the catalogue without pixelising the sky. The only operation applied to the data is the subtraction of the catalogue mean,
  \begin{equation} \label{eq:fj}
     f_j=\mathcal{D}^{\rm obs}_j-\Dcat,\qquad \Dcat=\frac{1}{\NF}\sum_{j\in \mathcal{S}}\mathcal{D}^{\rm obs}_j .
  \end{equation}
  We then construct the pseudo-$C_\ell$ estimator, subtracting the self-pair contribution to remove terms with $j=k$~\cite{2025JCAP...01..028W}. The resulting estimator and its bandpower form are
  \begin{equation} \label{eq:Ctilde}
    \tilde C_\ell=\frac{1}{4\pi}\sum_{\substack{j,k\in \mathcal{S}\\ j\neq k}}w_jw_kf_jf_kP_\ell({\nv}_j\cdot\nv_k),\qquad
    \tilde C_b=\sum_\ell B_{b\ell}\tilde C_\ell\equiv\sum_{\substack{j,k\in \mathcal{S}\\ j\neq k}}A^b_{jk}f_jf_k ,
  \end{equation}
  where $w_j$ are the catalogue weights, $P_\ell(x)$ is the Legendre polynomial, and $B_{b\ell}$ denotes the bandpower weights defined in Equation~\eqref{eq:bandpower}. The corresponding pair weights are
  \begin{equation} \label{eq:Ab}
    A^b_{jk}=\frac{w_jw_k}{4\pi}\sum_\ell B_{b\ell}P_\ell(\nv_j\cdot\nv_k).
  \end{equation}
 
  The survey mask couples multipoles, such that the expectation value of the pseudo-$C_\ell$ is related to the underlying angular power spectrum through
  \begin{equation} \label{eq:coupling}
    \big\langle\tilde C_\ell\big\rangle=\sum_{\ell'}M_{\ell\ell'}C_{\ell'},
  \end{equation}
  where $M_{\ell\ell'}$ is the mode-coupling matrix computed from the mask power spectrum. Subtracting self-pairs ensures the estimator is insensitive to white contributions from individual sightlines. 
  
  For a Gaussian field with $\langle f_jf_k\rangle=\Xi_{jk}$, the bandpower covariance can be evaluated exactly~\cite{Wolz2026},
  \begin{equation} \label{eq:exactcov}
    \mathrm{Cov}\big[\tilde C_b,\tilde C_{b'}\big]=2\sum_{\substack{j,k\in \mathcal{S}\\ j\neq k}}\sum_{\substack{l,m\in \mathcal{S}\\ l\neq m}}A^b_{jk}A^{b'}_{lm}\Xi_{jl}\Xi_{km}.
  \end{equation}
  Tests using Gaussian random fields sampled on the same mask show that this approximation \cite{2019JCAP...11..043G} reproduces the mean bandpowers $\langle \tilde{C}_b \rangle$ exactly, but can misestimate their variances by factors of up to $\sim50$ and their correlation coefficients by up to $\sim0.3$. We therefore compute the covariance directly using the exact-summation approach of Ref.~\cite{Wolz2026}.

  Subtracting the catalogue mean does not remove the redshift dependence of the mean DM--redshift relation. Writing $\mathcal{D}^{\rm obs}_j=\DMbar(z_j)+\mathcal{D}_j$, we can decompose the mean-subtracted field exactly as
  \begin{equation} \label{eq:split}
    f_j=\Delta_j+\epsilon_j,\qquad
    \Delta_j\equiv\DMbar(z_j)-\frac{1}{\NF}\sum_{k\in \mathcal{S}}\DMbar(z_k),\qquad
    \epsilon_j\equiv\mathcal{D}_j-\frac{1}{\NF}\sum_{k\in \mathcal{S}}\mathcal{D}_k.
  \end{equation}
  For an FRB without a measured redshift, $\Delta_j$ is unknown. We model it as an independent draw from the distribution of $\DMbar(z)$ over the redshift distribution $n(z)$, uncorrelated with the large-scale structure and with zero ensemble mean. It therefore contributes a white-noise term to the covariance, with variance
  \begin{equation} \label{eq:sigmaDelta}
    \sigma^2_\Delta=\int{\rm d}z\,n(z)\DMbar^2(z)-\left[\int{\rm d}z\,n(z)\DMbar(z)\right]^2,
  \end{equation}
  which enters $\Xi_{jk}$ in Equation~\eqref{eq:exactcov} together with the host scatter. Moreover, the product $\Delta_j\epsilon_k$ in the pseudo-$C_\ell$ estimator of Equation~\eqref{eq:Ctilde} couples the unknown mean-relation contribution to the DM fluctuations and can therefore contribute to the cross-covariance with host-identified FRBs. 
  
  Substituting Equation~\eqref{eq:split} into Equation~\eqref{eq:Ctilde} gives
  \begin{equation} \label{eq:Csplit}
    \tilde C_b=\sum_{j,k\in\mathcal{S}}A^b_{jk}\Delta_j\Delta_k+2\sum_{j,k\in\mathcal{S}}A^b_{jk}\Delta_j\epsilon_k+\sum_{j,k\in\mathcal{S}}A^b_{jk}\epsilon_j\epsilon_k .
  \end{equation}
  For fixed sightline positions and redshifts, the first term is deterministic. The covariance between the DM of a host-identified FRB $i$ and the bandpowers is consequently
  \begin{equation} \label{eq:crosscat}
    \Cov\big[\mathcal{D}_i,\tilde C_b\big]=\underbrace{\sum_{j,k\in\mathcal{S}}A^b_{jk}\big\langle\mathcal{D}_i\epsilon_j\epsilon_k\big\rangle}_{\text{three-point, Equation~\eqref{eq:crosscov_final}}}
    +\quad 2\sum_{j\in\mathcal{L},k\in\mathcal{S}}A^b_{jk}\,\Delta_j\langle\mathcal{D}_i\epsilon_k\rangle.
  \end{equation}
  The second term is expected to be small for two reasons. First, for random positions on the sky (even with a mask), the Legendre polynomial entering $A^b_{jk}$ averages to zero for $\ell \gg 1$ and $j\neq k$. The dominant contribution therefore comes from the auto-pair $j=i$, leaving a coherent term proportional to $\Delta_i$. Second, the term's magnitude is set by the relative cardinality of the sets $\mathcal{S}$ and $\mathcal{L}$. Typically, fewer than 10 per cent of FRBs are host-identified, further suppressing the contribution. This term largely averages out over host-identified FRBs (Section~\ref{ssec:sims.comparison}), and we include it in the exact Gaussian covariance used for the test in Section~\ref{ssec:forecast.chi2}.

\section{Validation against simulations} \label{sec:sims}
 \subsection{Numerical implementation} \label{ssec:sims.implementation}
  We evaluate all halo-model quantities in the Core Cosmology Library (\textsc{CCL}) \cite{Chisari2019}, defining halo masses using an overdensity threshold of $\Delta = 200$ relative to the critical density $\bar{\rho}_{\rm c}$. For the structural and statistical properties of haloes, we adopt the concentration--mass relation $c(M)$ of Ref.~\cite{Duffy2008}, the halo mass function $n(M)$ of Ref.~\cite{Tinker2008}, and the halo bias $b(M)$ of Ref.~\cite{Tinker2010}. As our fiducial cosmological model, we adopt a {\sl Planck} cosmology \cite{Planck2020}, with $\{\Omega_{\mathrm{c}}, \Omega_{\mathrm{b}}, h, n_{\mathrm{s}}, \sigma_8, \Sigma m_{\nu}\} = \{0.2607$, $0.04897$, $0.6766$, $0.9665$, $0.8102, 0.06\}$. We model the electron density profile using the baryonification framework of Ref.~\cite{Schneider2025}, described in Appendix~\ref{app:sec:profiles}. We take the FRB redshift distribution to be $n(z) \propto z^2 e^{-\alpha z}$ with $\alpha=3.5$, consistent with the parametrisation commonly adopted in recent analyses \cite[e.g.][]{Reischke2023, Sharma2025, Wayland2026}.

  To accelerate the cross-covariance computation, the electron bispectrum is precomputed on a grid in $(a, k_2, k_3, \phi)$ with $k_1 = \sqrt{k_2^2 + k_3^2 + 2 k_2 k_3 \cos\phi}$, consisting of 20 scale factors, 50 wavenumbers per axis, and 40 angles, interpolated thereafter. The connected non-Gaussian term is built with \texttt{halomodel\_Tk3D\_cNG} of \textsc{CCL} on 16 scale factors in $0 \leq z \leq 2$ and 48 logarithmically spaced wavenumbers in $10^{-4}\,{\rm Mpc}^{-1} \leq k \leq 10^3\,{\rm Mpc}^{-1}$, and projected with \texttt{angular\_cl\_cov\_cNG}.

 \subsection{Forward simulations} \label{ssec:sims.forward}
  To validate the analytical covariance, we generate mock FRB catalogues from forward simulations of the electron density field, so that we can measure both observables on the same realisations. Each catalogue contains the same $N_{\rm FRB}=10^4$ FRBs, with redshifts drawn from the survey redshift distribution and isotropic sky positions. We fix redshifts and positions across realisations so only the large-scale structure varies. We generate the electron density field with \textsc{glass} \cite{2016MNRAS.459.3693X,Tessore2023}, constructing 29 generated log-normal shells of equal comoving width between $z=0.01$ and $z=2$ (see \cite{2025OJAp....8E.102K} for details on the pipeline). We compute their input angular power spectra in the Limber approximation from the same halo model electron power spectrum as the theory, and we weight each shell by the DM kernel $(1+z)/E(z)$ within its redshift range. The maps have a \textsc{HEALPix} \cite{Gorski2005} resolution of $N_{\rm side}=512$ and are band-limited to $\ell_{\rm max}=3N_{\rm side}-1=1535$.

  \begin{figure}[ht]
      \centering
      \includegraphics[width=\linewidth]{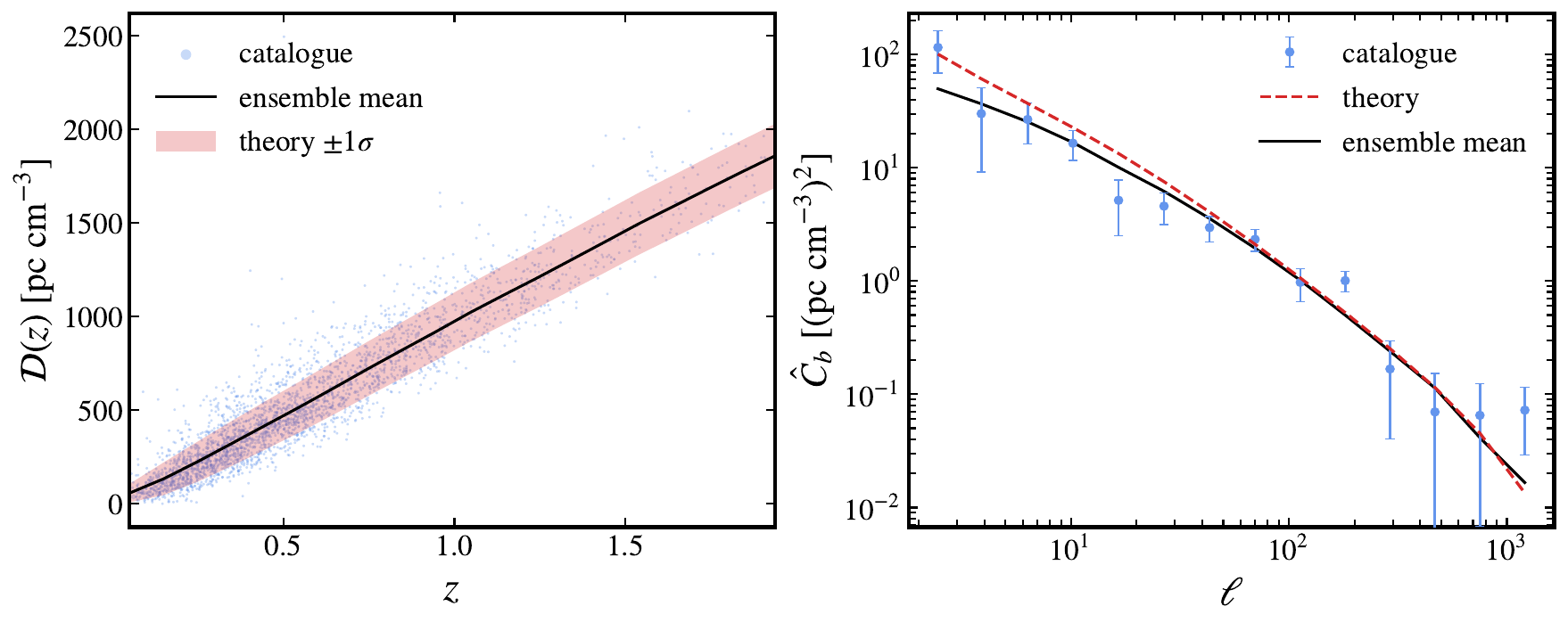}
      \caption{The two observables, measured on the same realisations. \emph{Left}: dispersion measure as a function of redshift for individual FRBs in one simulated catalogue. Points are the simulated FRBs, the solid line is the mean relation over the 5000 catalogues, and the shaded band is the theoretical $1\sigma$ scatter, $\sqrt{\mathrm{Var}(\mathcal{D}_i)}$. \emph{Right}: the decoupled angular power spectrum of the DM fluctuations, $\mathcal{D}_i - \langle \mathcal{D}_i \rangle$, measured from the same catalogue with the catalogue-based estimator of Equation~\eqref{eq:Ctilde} (points). Error bars show the $1\sigma$ scatter of the bandpowers of a single catalogue between realisations. The solid line is the mean over catalogues and the dashed line is the theoretical prediction. Both the theoretical prediction and the input spectra of the simulations use the Limber approximation.}
      \label{fig:dmz}
  \end{figure}

  The DM of each FRB is calculated as the sum of the contributions from all foreground shells, plus a linearly weighted contribution from the shell containing the FRB, and the mean Macquart relation. The chosen resolution is expected to be sufficient for our forecast, as the number of FRBs considered is not expected to provide sensitivity to multipoles $\ell \gtrsim 10^3$. However, due to this band limitation, the variance of each shell map is lower than that implied by its input power spectrum. To account for this, we measure the reduced map variance 
  \begin{equation}
      \tilde{\sigma}^2 = \mathrm{Var}(\delta_{\rm map}),  
   \end{equation}
   compute the deficit $\Delta\sigma^2 = \sigma^2 - \tilde{\sigma}^2$ with 
  \begin{equation}
      \sigma^2 = (4\pi)^{-1} \sum_\ell (2\ell+1)\, C_\ell,
  \end{equation}
  and restore it by adding an isotropic Gaussian component $\eta(\nv) \sim \mathcal{N}(0,\Delta\sigma^2)$, uncorrelated between pixels, to each shell map. This reduces the need for large $N_\mathrm{side}$ while restoring the total variance expected from the input spectrum. A small caveat is that this approach can, in principle, produce a completely correlated signal if FRBs fall into the same pixel. In our catalogues, only 30 of the $10^4$ FRBs share a pixel with another, so this effect is negligible.
  
  The left-hand panel of Figure~\ref{fig:dmz} shows the DMs of one representative catalogue against the mean Macquart relation and the $1\sigma$ band predicted by Equation~\eqref{eq:covDD_legendre}. The simulated FRBs remain centred on the mean relation across the full redshift range, and the standard deviation of the DM of each FRB over the 5000 catalogues is consistent with $\sqrt{\mathrm{Var}(\mathcal{D}_i)}$ to within $1$--$3$ per cent. Also note that the shaded band indicates a symmetric scatter, while the simulation scatter is intrinsically non-Gaussian \cite{2021ApJ...906...49Z,2021MNRAS.505.5356B,2025arXiv250707090K,2026OJAp....961522T}, as it sums log-normal random variables with changing variance.

 \subsection{Comparison of theory and simulation covariances} \label{ssec:sims.comparison}
  \begin{figure}[ht]
      \centering
      \includegraphics[width=\linewidth]{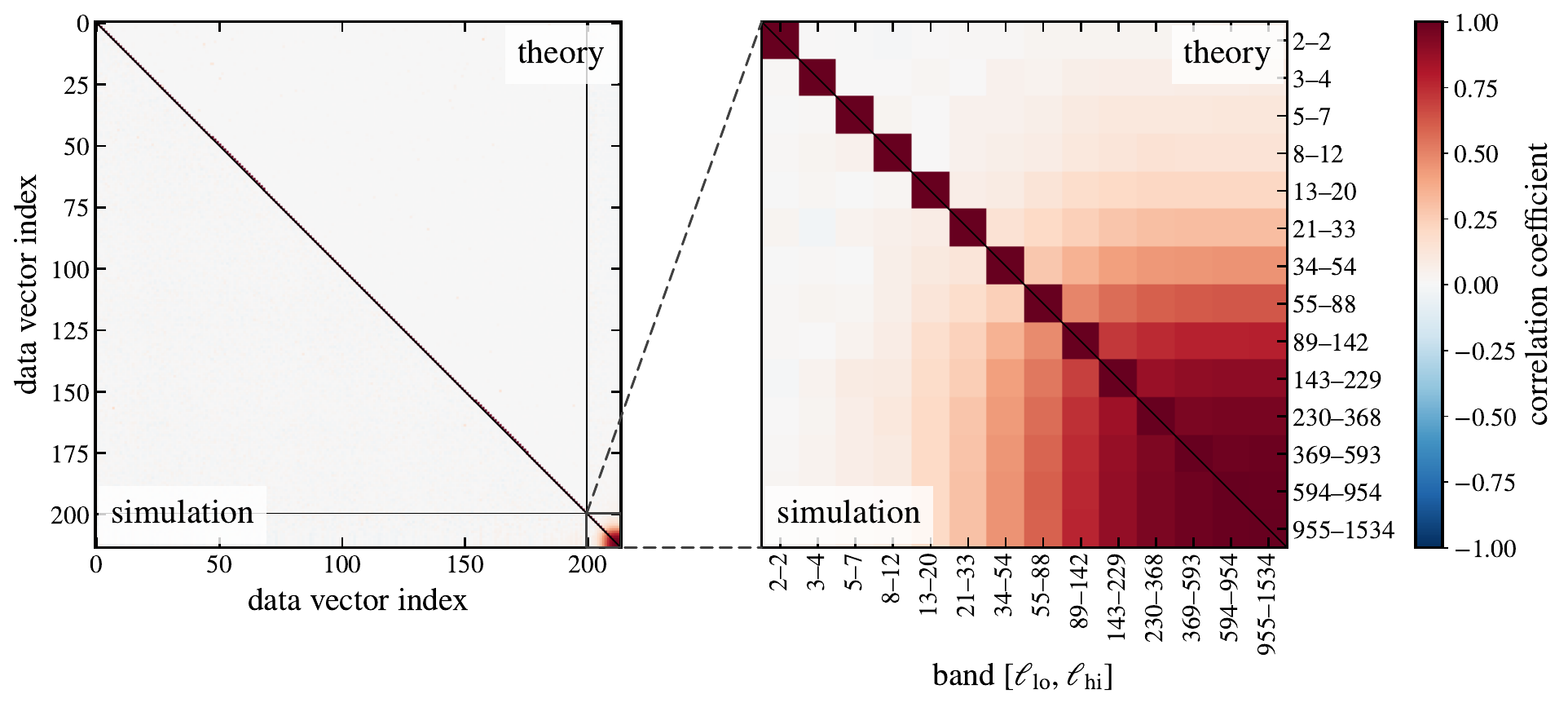}
      \caption{Correlation coefficients of the full data vector (\emph{left}) and of the binned DM pseudo-power spectrum (\emph{right}) for $10^4$ FRBs at fixed positions and redshifts. In both panels, the upper triangle shows the analytical covariance and the lower triangle shows the sample covariance of 5000 log-normal forward simulations. Bands are labelled by their range $[\ell_{\rm lo}, \ell_{\rm hi}]$. In the left-hand panel, the data vector runs over the DMs of the 200 FRBs (indices 0--199, ordered by redshift) followed by the 14 bandpowers (indices 200--213), with black lines separating the two blocks. The corresponding correlations for the redshift-binned DMs used in the forecast are shown in Figure~\ref{fig:corr_binned}.}
      \label{fig:corr}
  \end{figure}

  The right-hand panel of Figure~\ref{fig:dmz} shows the angular power spectrum of the same catalogue. To compare it directly with the theoretical prediction, we measure it with the catalogue-based estimator of Equation~\eqref{eq:Ctilde}, which removes the self-pair terms, and invert the mode coupling of the catalogue mask~\cite{Alonso2019, 2025JCAP...01..028W}. We subtract the ensemble-mean DM of each FRB, $\langle\mathcal{D}_i\rangle$, rather than the catalogue mean, so the spectrum contains only the fluctuations $\epsilon_j$ of Equation~\eqref{eq:split}. The decoupled spectrum agrees with the theory to within approximately 5 per cent for $\ell \gtrsim 70$.

  For each catalogue, we measure the pseudo-$C_\ell$ of the pixelised DM fluctuation, $\mathcal{D}_i - \bar{\mathcal{D}}_{\rm cat}$, as in Equation~\eqref{eq:fj}, at $N_{\rm side}=512$, without subtracting the self-pairs, and bin it into 14 logarithmically spaced bands $[\ell_{\rm lo},\ell_{\rm hi})$ between $\ell=2$ and $\ell=1535$. Here, we apply no mode-coupling deconvolution to either the theory or the simulation, since the covariance of the coupled bandpowers can be evaluated exactly (Section~\ref{ssec:theory.catalogue_pcl}). The joint data vector,
  \begin{equation} \label{eq:datavector}
      \mathbf{d} = \left[\mathcal{D}_{1}^{\rm obs}, \dots, \mathcal{D}_{200}^{\rm obs}, \tilde{C}_{b=1}, \dots, \tilde{C}_{b=14}\right],
  \end{equation}
  consists of the DMs of 200 FRBs, ordered by redshift (this can be thought of as the sub-sample with a spectroscopically confirmed host galaxy), and the 14 bandpowers, and we estimate its sample covariance from $N_{\rm sim}=5000$ catalogues. In Section~\ref{ssec:sims.comparison}, we also consider a compressed version of this data vector, in which the DMs are averaged in redshift bins, as in the forecast of Section~\ref{sec:forecast}.

  The left-hand panel of Figure~\ref{fig:corr} shows the correlation coefficients of the full data vector. The DMs of the 200 FRBs are nearly uncorrelated, as expected for sightlines that are mostly widely separated on the sky, and their correlation with the bandpowers is weak in both theory and simulation. The right-hand panel zooms in on the bandpower block, comparing the correlation coefficients from theory and simulations. The theory reproduces the simulated correlation structure across all bands: bands with $\ell \lesssim 20$ are nearly uncorrelated, correlations build up between $\ell \approx 20$--$50$, and bands with $\ell \gtrsim 90$ are almost fully correlated.

  \begin{figure}[ht]
      \centering
      \includegraphics[width=0.9\linewidth]{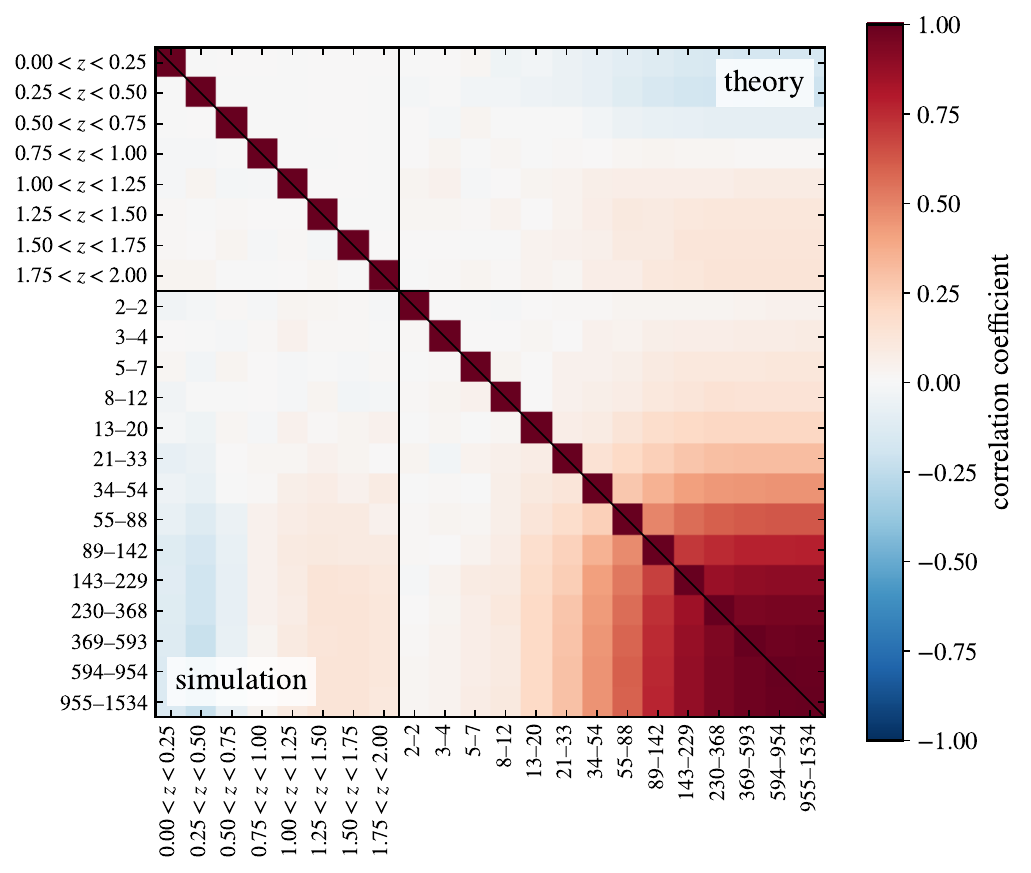}
      \caption{As the left-hand panel of Figure~\ref{fig:corr}, for the compressed data vector: the mean DMs of the 1444 host-identified FRBs in redshift bins of width $\Delta z = 0.25$ (labelled by redshift range), followed by the 14 bandpowers. The upper triangle shows the analytical covariance propagated through the binning, and the lower triangle shows the sample covariance of the 5000 simulations.}
      \label{fig:corr_binned}
  \end{figure}

  In practice, the DMs of the host-identified FRBs can be compressed before they enter the analysis, and we do so in the forecast of Section~\ref{sec:forecast}. We order them by redshift, assign them to bins of width $\Delta z = 0.25$ over $0 < z < 2$, and average their DMs within each bin. Since this is a linear transformation of the data vector, the covariance of the binned means and their cross-covariance with the bandpowers follow exactly from those of the individual DMs. Averaging within a bin discards the variation of the mean relation across the bin, so some information is lost, but this loss is small for bins over which $\bar{\mathcal{D}}(z)$ varies slowly. 
  
  Figure~\ref{fig:corr_binned} shows the correlation coefficients of the compressed data vector, in which we designate 1444 of the $10^4$ simulated FRBs as host-identified, matching the number used in the forecast of Section~\ref{sec:forecast}. Again, the theory and simulation agree well: the binned DMs remain nearly uncorrelated with each other because, on the full sky, they share no large-scale modes for $\ell \geq 1$. Their correlations with the bandpowers, however, are clearly visible, reaching $|r| = 0.21$ in the theory and 0.22 in the simulations. These values are substantially larger than the correlation of individual DMs, which are typically $|r_i| \approx 0.01$. Indeed, 99 per cent of the entries in Figure~\ref{fig:corr} satisfy $|r_i|<0.04$, with maximum values of 0.06 in the theory and 0.08 in the simulations. The additional scatter in the latter is consistent with the sampling noise of $N_{\rm sim}^{-1/2} \approx 0.014$ per entry. 
  
\begin{figure}[ht]
      \centering
      \includegraphics[width=0.7\linewidth]{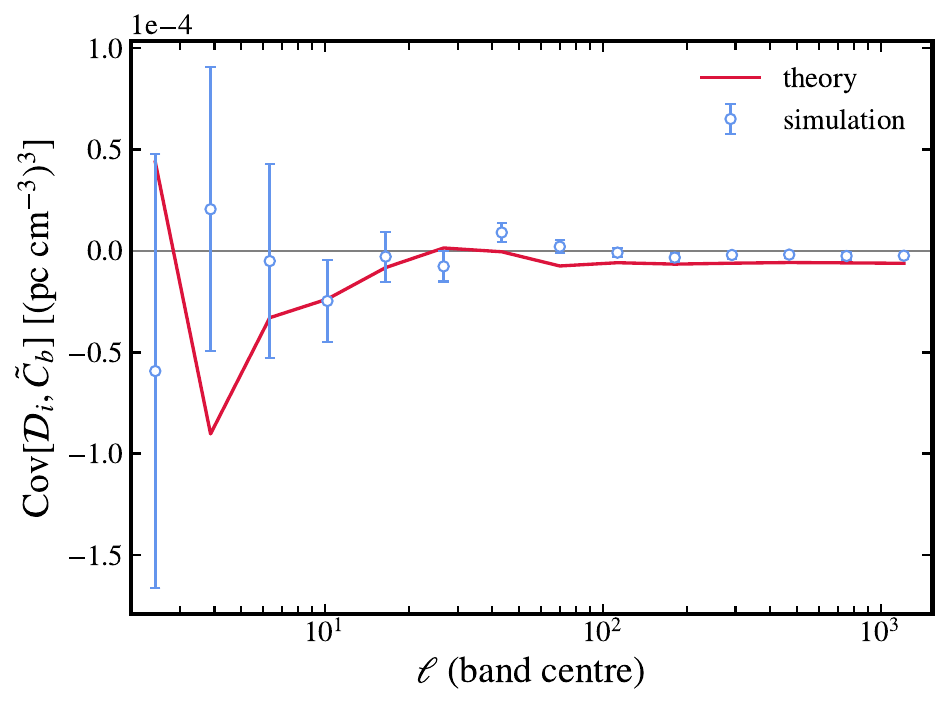}
      \caption{The $\mathcal{D}$--$\tilde{C}_\ell$ block of the covariance, averaged over the 200 FRBs of the data vector, in units of $({\mathrm{pc\,cm}}^{-3})^3$. Points show the sample covariance of 5000 log-normal forward simulations with standard errors over the 5000 realisations, and the line shows the theoretical prediction.}
      \label{fig:cross_entries}
  \end{figure}

  The difference arises from the second term in Equation~\eqref{eq:crosscat}. For $\ell \gtrsim 200$, where the binned correlation is largest, a term linear in $\Delta_i$ accounts for 60--77 per cent of the variation of $\mathrm{Cov}[\mathcal{D}_i,\tilde{C}_b]$ between FRBs, causing the correlation to change sign with redshift. Within a redshift bin, however, the $N_b$ FRBs share the same sign of $\Delta_i$. Averaging over their DMs therefore reduces the variance of the mean DM by approximately a factor of $N_b$, while leaving its covariance with the bandpowers unchanged, so that $r_{\rm bin} \approx N_{b}^{1/2} \langle r_i \rangle_b$. In the bin $0.25 < z < 0.50$, for example, $N_b = 458$ and $\langle r_i \rangle_b = -0.01$, which gives $r_{\rm bin}=-0.21$. At higher redshifts, $\langle r_i \rangle_b$ is larger, reaching 0.03, but the bins contain only 19--62 FRBs, resulting in $|r_{\rm bin}| \approx 0.12$--$0.24$. The binned means of the 200 FRBs of Equation~\eqref{eq:datavector} follow the same relation. In contrast, averaging over all host-identified FRBs cancels this contribution because the $\Delta_i$ sum to zero, leaving $|r|<0.005$, as shown in Figure~\ref{fig:cross_entries}.

  Figure~\ref{fig:cross_entries} shows the $\mathcal{D}$--$\tilde{C}_\ell$ block itself, averaged over the 200 FRBs of the data vector. The theoretical prediction in Equation~\eqref{eq:crosscat} and the simulation sample covariance agree within errors of the sample covariance across the bandpowers. Both are small: expressed as a correlation coefficient, each FRB-averaged entry satisfies $|r|<0.005$. For $N_{\rm sim} = 5000$ realisations, the sampling error on a single correlation coefficient estimated from the simulations is $N_{\rm sim}^{-1/2} \simeq 0.014$, and on the average over the 200 FRBs approximately $10^{-3}$, so the averaged entries are consistent with zero. Normalising the averaged cross-covariance by the corresponding standard deviations gives correlation coefficients of order $10^{-4}$--$3\times10^{-3}$, of the same order as the full-sky value in Figure~\ref{fig:fsky_scaling}.

 \subsection{Dependence on the survey footprint} \label{ssec:sims.footprint}
\begin{figure}
      \centering
      \includegraphics[width=0.7\linewidth]{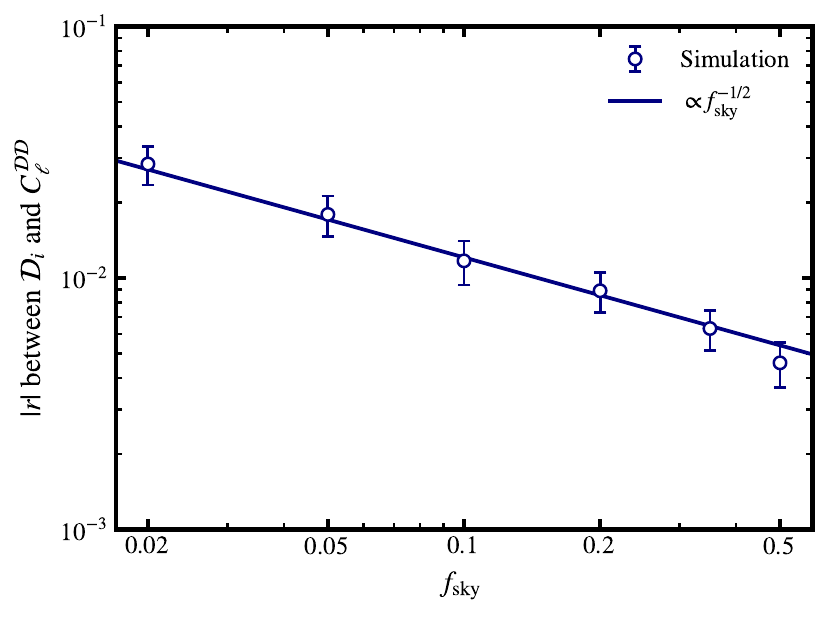}
      \caption{Correlation coefficient between an individual FRB's DM and the binned DM angular power spectrum as a function of the sky fraction, $f_{\rm sky}$, of the survey footprint. We obtain measurements from 500 forward simulations. Error bars show the sampling uncertainty estimated from the 500 realisations, rescaled by $\sigma(\bar{\mathcal{D}})/\sigma(\mathcal{D}_i)$, which varies with the footprint. The solid line shows the predicted $f_{\rm sky}^{-1/2}$ scaling from Equation~\eqref{eq:r_bound_window}.}
      \label{fig:fsky_scaling}
  \end{figure}

  Equation~\eqref{eq:r_bound_window} predicts that the cross-correlation magnitude increases as the footprint decreases, with $|r| \propto f_{\rm sky}^{-1/2}$. This scaling follows from the inverse-area dependence of the window variance, $\sigma_\mathrm{b}^2 \propto \Omega_{\rm s}^{-1}$, in Equation~\eqref{eq:sigmab}. Here, we test the prediction directly using the forward simulations. Figure~\ref{fig:fsky_scaling} shows the resulting footprint dependence. The correlation between an individual FRB's DM and the bandpowers follows $|r| \propto f_{\rm sky}^{-0.54 \pm 0.07}$, consistent with the prediction of Equation~\eqref{eq:r_bound_window}. On very large scales, we see edge effects because naive scaling with $f_\mathrm{sky}$ requires an infinitely extended sky.
  
  Although the cross-correlation grows as the survey footprint shrinks, it remains negligible for realistic FRB surveys. Extrapolating the measured scaling, $|r| \simeq r_1 f_{\rm sky}^{-1/2}$ with $r_1 \approx 4 \times 10^{-3}$ for a full-sky survey, the correlation coefficient reaches $|r| \approx 0.06$ at $f_{\rm sky} \approx 4 \times 10^{-3}$. This corresponds to a footprint of approximately $150\,{\rm deg}^2$, which is far smaller than that of any planned wide-field FRB survey.

\section{Forecast} \label{sec:forecast}
 \subsection{Does the cross-covariance matter for inference?} \label{ssec:forecast.chi2}
  A correlation coefficient below $|r|<0.005$ for the FRB-averaged cross-covariance of the individual DMs is not, by itself, sufficient justification\footnote{It is, however, a clear hint.} for dropping the cross-covariance: what matters is its impact on parameter inference, rather than its size relative to the diagonal. This is especially true for the compressed data vector, whose correlation coefficients reach $|r| \approx 0.2$. We therefore test the cross-covariance at the level where it enters the analysis, following the validation strategy of Ref.~\cite{Wolz2026}, to determine whether the Macquart relation and the angular power spectrum can be treated as independent probes. Specifically, we draw $10^5$ mock data vectors from the full covariance $\mathsf{C}$ and analyse each mock twice: once using $\mathsf{C}$, and once using the block-diagonal covariance $\mathsf{C}_0$ obtained by setting the cross-covariance to zero. Comparing the resulting $\chi^2$ values, and the corresponding probabilities to exceed (PTEs), directly quantifies the effect of neglecting the cross-covariance.

\begin{figure}[ht]
      \centering
      \includegraphics[width=\linewidth]{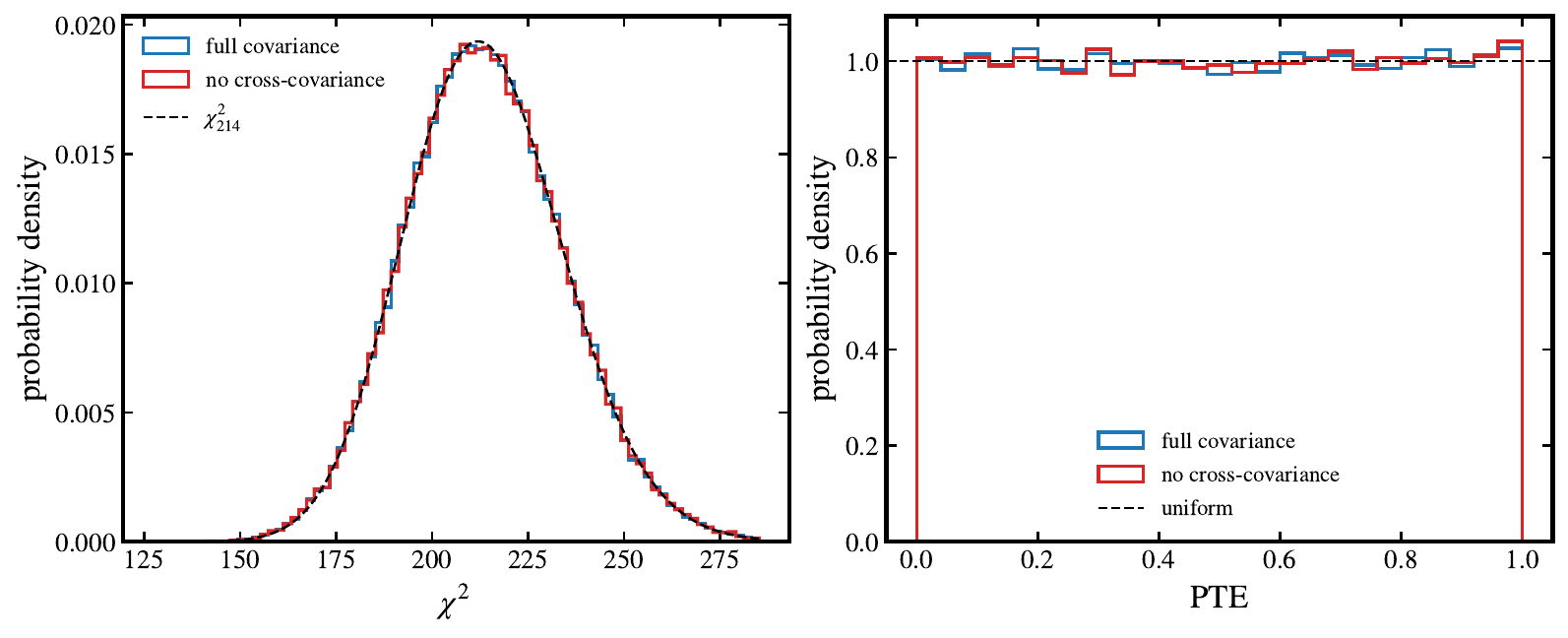}
      \caption{Distribution of $\chi^2$ (\emph{left}) and the probability to exceed (\emph{right}) for $10^5$ mock data vectors drawn from the full covariance $\mathsf{C}$, and analysed with $\mathsf{C}$ (blue) and with the block-diagonal $\mathsf{C}_0$ obtained by setting the cross-covariance to zero (red). The $\chi^2$ distribution for $n=214$ degrees of freedom, $\chi^2_{214}$, is shown for reference.}
      \label{fig:chi2_test}
  \end{figure}

  The two distributions are nearly indistinguishable, as shown in Figure~\ref{fig:chi2_test}. The standard deviation of $\chi^2$ changes from 20.80 to 20.84, corresponding to a 0.2 per cent shift. The Kolmogorov-Smirnov (KS) statistics for the two PTE distributions against the uniform distribution are 0.26 and 0.29, respectively; neither is significant for $10^5$ samples. Additionally, the fraction of mocks for which the acceptance decision at 95 per cent confidence differs between the two treatments is 0.5 per cent.

  We repeat the test for the compressed data vector of Figure~\ref{fig:corr_binned}. Despite its larger correlation coefficients, neglecting the cross-covariance changes the width of the $\chi^2$ distribution by only 0.8 per cent and the acceptance decision at 95 per cent confidence for 1.0 per cent of the realisations. For the same 1444 FRBs treated individually, the width of the $\chi^2$ distribution changes by 0.2 per cent, as for the 200 FRBs above. The effect grows with the number of FRBs per bin, but remains at the per cent level for the forecast considered here. The test is conservative in the respect that the simulations hold the redshifts of all $10^4$ FRBs fixed, so that $\Delta_j$ is deterministic for every burst, whereas for real data only the host-identified FRBs $j \in L$ contribute to the second term of Equation~\eqref{eq:crosscat}.

 \subsection{Survey specifications} \label{ssec:forecast.survey}
  Using the covariance validated above, we forecast the joint constraints from a Stage-IV weak lensing (WL) survey and an FRB catalogue,  extending the methodology of Ref.~\cite{Wayland2026}. We model the WL survey as LSST-like, following the specifications of the LSST Dark Energy Science Collaboration Science Requirements Document~\cite{Mandelbaum2018} for the 10-year WL sample. We adopt the source number density, redshift distribution, and photometric redshift uncertainty as described in Ref.~\cite{Leonard2023}, and divide the source sample into the same five equal-density redshift bins. The resulting cosmic shear data vector comprises all the auto- and cross-power spectra between pairs of redshift bins.
  
  The FRB sample is assumed to have a number density of $0.5\,\mathrm{deg}^{-2}$ over a sky fraction of $f_{\rm sky}=0.7$, with a host galaxy DM standard deviation of $\sigma_{\rm host,0}=90\,{\rm pc\,cm}^{-3}$, motivated by current experiments such as CHIME/FRB~\cite{Rafiei2021,2024ApJ...973..151K}. We adopt the same FRB redshift distribution, $n(z) \propto z^2 e^{-\alpha z}$, used throughout this work, and assume that $10$ per cent of the FRBs, 1444 of the 14439 in the footprint, have a host galaxy with a spectroscopic redshift. In the forecast, we allow for uncertainty in this distribution by modelling it as $n(z) \propto (z+\delta z)^2 e^{-\alpha z}$, with $\alpha$ and $\delta z$ free. The total data vector then consists of the FRB DM auto-power spectrum, its cross-power spectra with each of the five WL bins, the fifteen WL auto- and cross-power spectra, and the mean DM of the host-identified FRB subsample over the redshift range $0<z<2$. We impose scale cuts of $\ell_{\rm max}^{\mathcal{DD}} = 500$ for the DM auto-power spectrum and $\ell_{\rm max}^{\mathcal{D}\gamma} = 1000$ for the DM--WL cross-power spectra.

  As described in Section~\ref{ssec:sims.comparison}, we compress the Macquart data into the mean DMs of the host-identified FRBs in 8 redshift bins of width $\Delta z = 0.25$ over $0 < z < 2$, as in Figure~\ref{fig:corr_binned}. Following Section~\ref{ssec:forecast.chi2}, we neglect the cross-covariance between these binned means and the angular power spectra.

 \subsection{Joint constraints}
 \label{ssec:forecast.joint_constraints}
  We sample the joint posterior of the WL and FRB data vector described above, adopting the covariance validated in Section~\ref{sec:sims}. We marginalise over the cosmological parameters $\{\omega_{\rm b}, \Omega_{\rm m}, \sigma_8, h\}$, the baryonification parameters $\{\log_{10} M_{\rm c}, \delta\}$, the mean host contribution ${\rm DM}_{\rm host}$, and the parameters $\alpha$ and $\delta z$ of the FRB redshift distribution. We report the derived combination $S_8 \equiv \sigma_8 \sqrt{\Omega_{\rm m}/0.3}$. The baryonic suppression of the matter power spectrum is computed with the \textsc{BCemu} emulator~\cite{Giri2021} applied to the non-linear matter power spectrum, and the electron distribution follows the corresponding baryonification gas profile~\cite{Schneider2019}.

  Here, $M_{\rm c}$ is the characteristic halo mass below which gas is efficiently expelled from haloes, and $\delta$ sets the slope of the gas profile beyond the ejection radius. The remaining parameters of the model are held fixed (see Appendix~\ref{app:sec:profiles}), and we adopt uniform priors $\log_{10} M_{\rm c} \in [11,15]$ and $\delta \in [1,11]$. The host-identified subsample has spectroscopic redshifts, so we measure the distribution of sources within each Macquart bin rather than model it. We therefore hold $\alpha$ and $\delta z$ at their fiducial values in the Macquart block while allowing them to vary in the angular power spectra, where the redshifts of individual bursts are unknown.

  \begin{figure}[ht]
      \centering
      \includegraphics[width=\linewidth]{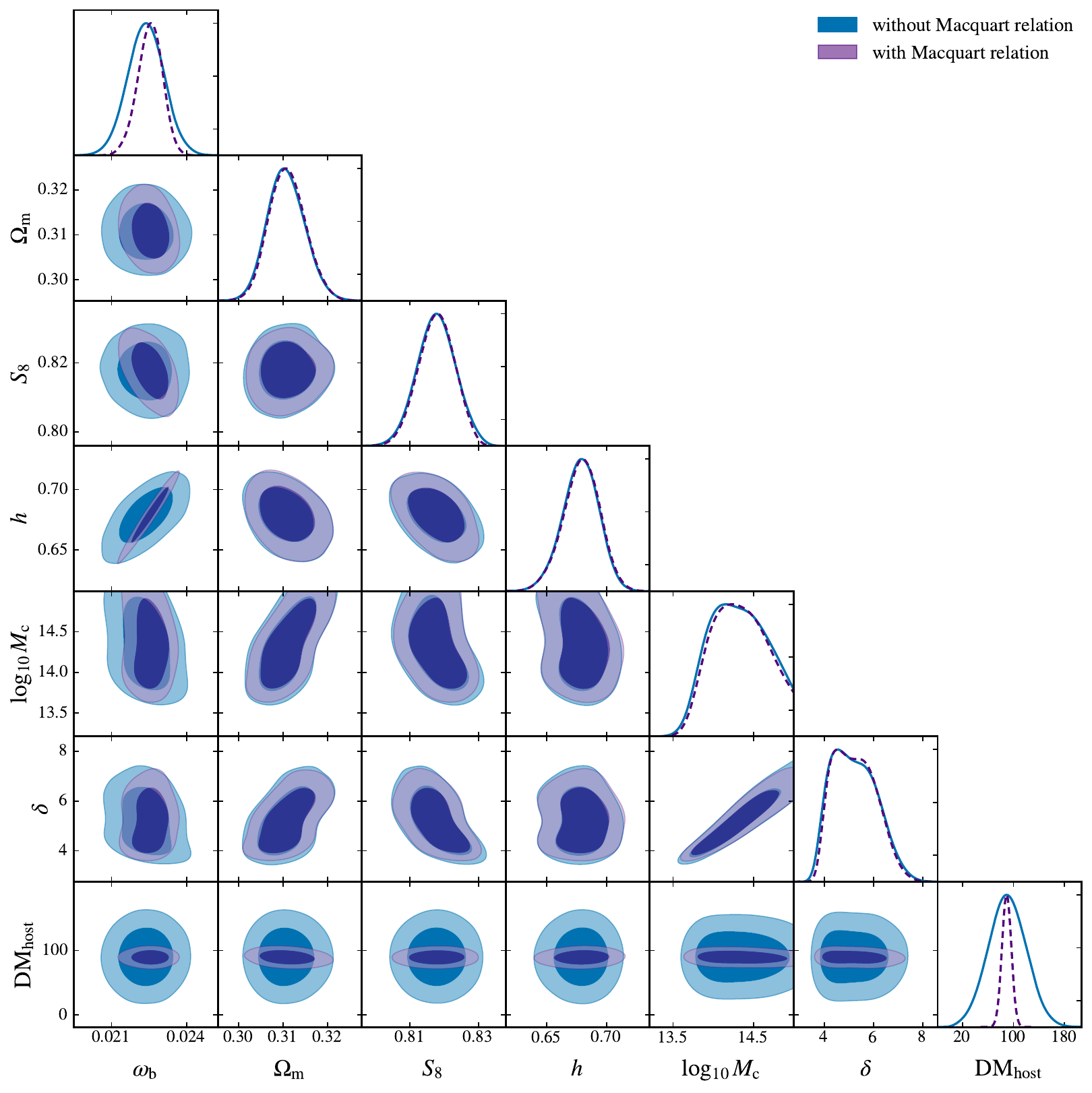}
      \caption{The marginalised posteriors for the joint LSST-like WL and FRB forecast, without the Macquart relation in the data vector (blue) and with it included (purple). Contours enclose the $68$ and $95$ per cent credible regions. Both analyses use the same angular power spectra and marginalise over ${\rm DM}_{\rm host}$, $\alpha$, and $\delta z$, with $\alpha$ and $\delta z$ held fixed at their fiducial values within the Macquart block.}
      \label{fig:corner}
  \end{figure}

  \begin{table}[ht]
      \centering
      \begin{tabular}{lll}
        \hline
        Parameter & without Macquart & with Macquart \\
        \hline
        \hline
        $\omega_{\rm b}$ & $0.02238 \pm 0.00072$ & $0.02254 \pm 0.00049$ \\
        $\Omega_{\rm m}$ & $0.3107 \pm 0.0041$ & $0.3109 \pm 0.0041$ \\
        $S_8$ & $0.8177 \pm 0.0056$ & $0.8177 \pm 0.0053$ \\
        $h$ & $0.678 \pm 0.015$ & $0.679 \pm 0.015$ \\
        $\log_{10} M_{\rm c}$ & $>13.78$ & $>13.81$ \\
        $\delta$ & $5.21 \pm 0.94$ & $5.23 \pm 0.88$ \\
        ${\rm DM}_{\rm host} \,[{\rm pc\,cm}^{-3}]$ & $90.3 \pm 29.9$ & $89.3 \pm 7.3$ \\
        \hline
      \end{tabular}
      \caption{The marginalised means and $68$ per cent credible intervals from the joint WL and FRB forecast, without and with the Macquart relation included in the data vector. The posterior of $\log_{10} M_{\rm c}$ extends to the upper edge of its prior, so we quote the $95$ per cent lower limit.}
      \label{tab:constraints}
  \end{table}

  Figure~\ref{fig:corner} shows the posterior distributions with and without the Macquart relation included in the data vector, and Table~\ref{tab:constraints} lists the corresponding marginalised constraints. The two analyses share the same angular power spectra and differ only in whether the binned mean DM measurements of the host-identified subsample are included.
  
  We leave the physical baryon density $\omega_{\rm b} = \Omega_{\rm b} h^2$ free rather than fixing it. The amplitude of the Macquart relation scales as $\Omega_{\rm b} H_0^2$, while the line-of-sight distance integral contributes a factor of $H_0^{-1}$. Consequently, the mean DM measures the combination $\Omega_{\rm b} H_0 = \omega_{\rm b}/h$ and not either parameter separately \cite{2019MNRAS.484.1637J,hagstotz_new_2022}. Fixing $\omega_{\rm b}$ would therefore convert this measurement into an apparent constraint on $h$ that is imposed by the prior assumption rather than independently determined by data. The $\omega_{\rm b}$--$h$ panel of Figure~\ref{fig:corner} shows the consequence directly: including the Macquart relation collapses the contour onto a narrow, positively inclined band, which is the locus of constant $\omega_{\rm b}/h$ anticipated above, while leaving the extent along the band almost unchanged. The Macquart relation improves $\sigma(\omega_{\rm b})$ by $32$ per cent, from $7.2 \times 10^{-4}$ to $4.9 \times 10^{-4}$, while leaving $h$ unchanged at the per cent level, as expected if only the combination is measured.

  The effect of the Macquart relation is concentrated in the host contribution. The uncertainty on ${\rm DM}_{\rm host}$ decreases from $29.9$ to $7.3\,{\rm pc\,cm}^{-3}$, a factor of $4.1$. This is expected because the mean DM is the only observable in the data vector sensitive to a redshift-dependent additive offset, since such an offset contributes no angular fluctuation power and therefore leaves $C_\ell^{\mathcal{DD}}$ unchanged. The constraints on $\Omega_{\rm m}$ and $S_8$ change by at most $5$ per cent, as expected, since the WL and DM angular power spectra already constrain these parameters much more tightly than the binned mean DMs.

  The constraints on the baryonification parameters improve by approximately $5$ per cent, with the standard deviation of the posterior on $\log_{10} M_{\rm c}$ falling from $0.37$ to $0.36$ and $\sigma(\delta)$ from $0.94$ to $0.88$. This improvement is indirect, since the mean Macquart relation has no direct sensitivity to feedback in our model. Baryonification redistributes gas within and around haloes, modifying the electron power spectrum on all but the largest scales, but the modelled mean electron column density to a given redshift is set by $\Omega_{\rm b}$ alone. As a result, the modelled mean DM is independent of $\log_{10} M_{\rm c}$ and $\delta$ by construction.

  Instead, the Macquart relation constrains ${\rm DM}_{\rm host}$. Tightening the host constraint by a factor of $4.1$ reduces the freedom associated with this nuisance parameter, allowing the feedback constraints to improve indirectly. The similar fractional improvements in $\log_{10} M_{\rm c}$ and $\delta$ are consistent with both parameters benefiting from reduced shared degeneracy, rather than from independent information on either parameter. A strong degeneracy between the two baryonification parameters limits the gain, visible as the positively inclined $\log_{10} M_{\rm c}$--$\delta$ contour in Figure~\ref{fig:corner}. The data constrain $\log_{10} M_{\rm c}$ from below but not from above, with its posterior extending to the upper prior boundary in both analyses.

 \subsection{Redshift self-calibration from a joint analysis} \label{sec:joint_relevance}
  \begin{figure}[]
      \centering
      \includegraphics[width=\textwidth]{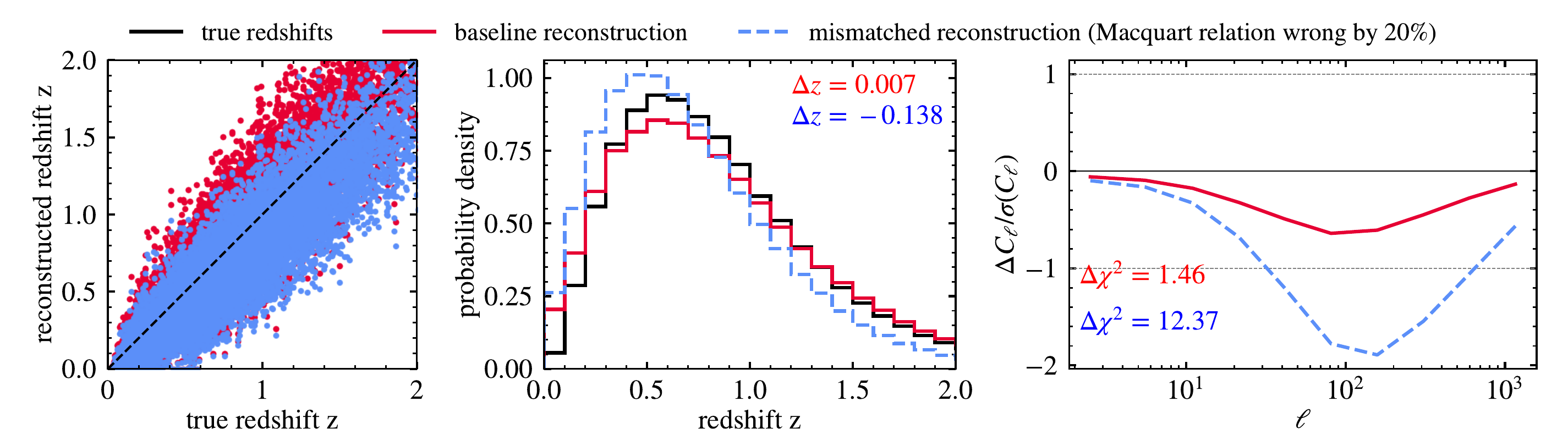}
      \caption{Reconstruction of the redshift distribution for FRBs with an unidentified host. The true redshifts are shown in black, the baseline reconstruction with an unbiased Macquart relation in red, and the mismatched reconstruction (Macquart relation wrong by 20\%) in blue. \emph{Left}: reconstructed redshift against true redshift. The dashed line indicates the identity. \emph{Middle}: corresponding redshift distributions and their shift of the mean $\Delta z$. \emph{Right}: Bias in the estimated $C_\ell$ in units of its measurement error for $5\times10^5$ FRBs. The $\Delta\chi^2$ shows the cumulative difference.  }
      \label{fig:redshift_reconstruction}
  \end{figure}
  In Section~\ref{ssec:forecast.chi2}, we showed that the cross-covariance between the Macquart relation and the DM angular power spectrum can be neglected. Furthermore, we showed that adding the Macquart relation does not significantly improve estimates of the feedback parameters, largely because the two data sets have different statistical power: the angular power spectrum is estimated from the full catalogue, most of which consists of FRBs without an identified host. However, this still matters. To interpret the theoretical prediction of the angular power spectrum, we need the FRB redshift distribution entering the kernel $W_\mathcal{D}$ in Equation~\eqref{eq:kernel}. For an FRB without an identified host, the only redshift information is its observed DM, so we must infer the redshift distribution of this sample by inverting the mean DM--redshift relation. The Macquart relation therefore enters the model of every angular power spectrum involving the DM field, and any error in it propagates into the inferred $n(z)$ and consequently into the predicted $C_\ell^{\mathcal{D}\mathcal{D}}$.

  We illustrate this in Figure~\ref{fig:redshift_reconstruction}. For each FRB without an identified host, we estimate its redshift $\hat{z}_i$ by solving
  \begin{equation} \label{eq:DM_i_obs}
      \mathcal{D}^\mathrm{obs}_i = \bar{\mathcal{D}}(\hat{z}_i)
      + \mathrm{DM}_\mathrm{host}/(1+\hat{z}_i),
  \end{equation}
  and construct the redshift distribution from the histogram of the $\hat{z}_i$. We do this directly in our log-normal simulations and compare a baseline reconstruction, where the assumed mean relation matches the true one, with a mismatched reconstruction in which the assumed mean relation overestimates the true one by 20 per cent for illustrative purposes. In the baseline case, the scatter about the mean relation, caused by the large-scale structure and the host contribution, smears the individual redshift estimates considerably (left panel), but the reconstructed distribution remains close to the true one, and its mean is shifted by only $\Delta z = 0.007$ (middle panel). In the mismatched case, every burst is assigned a systematically lower redshift, and the distribution mean shifts by $\Delta z = -0.138$. 

  The right-hand panel of Figure~\ref{fig:redshift_reconstruction} shows the consequence for the angular power spectrum. We compute $C_\ell^{\mathcal{D}\mathcal{D}}$ from each reconstructed redshift distribution and compare it with the prediction from the true distribution, in units of the measurement uncertainty for a catalogue of $5\times10^5$ FRBs. The baseline reconstruction biases the power spectrum by less than $1\sigma$ in every band, with a cumulative $\Delta\chi^2 = 1.46$, reflecting broadening of the reconstructed distribution rather than a shift in its mean. The mismatched reconstruction under-predicts the power spectrum by up to $2\sigma$ in individual bands at $\ell \simeq 100$--$300$, accumulating to a large $\Delta\chi^2$. Unless the redshift distribution is modelled with the correct mean relation, this bias would be absorbed by the parameters that control the amplitude and shape of the spectrum on these scales, including the baryonification parameters.

  \begin{figure}[]
      \centering
      \includegraphics[width=\linewidth]{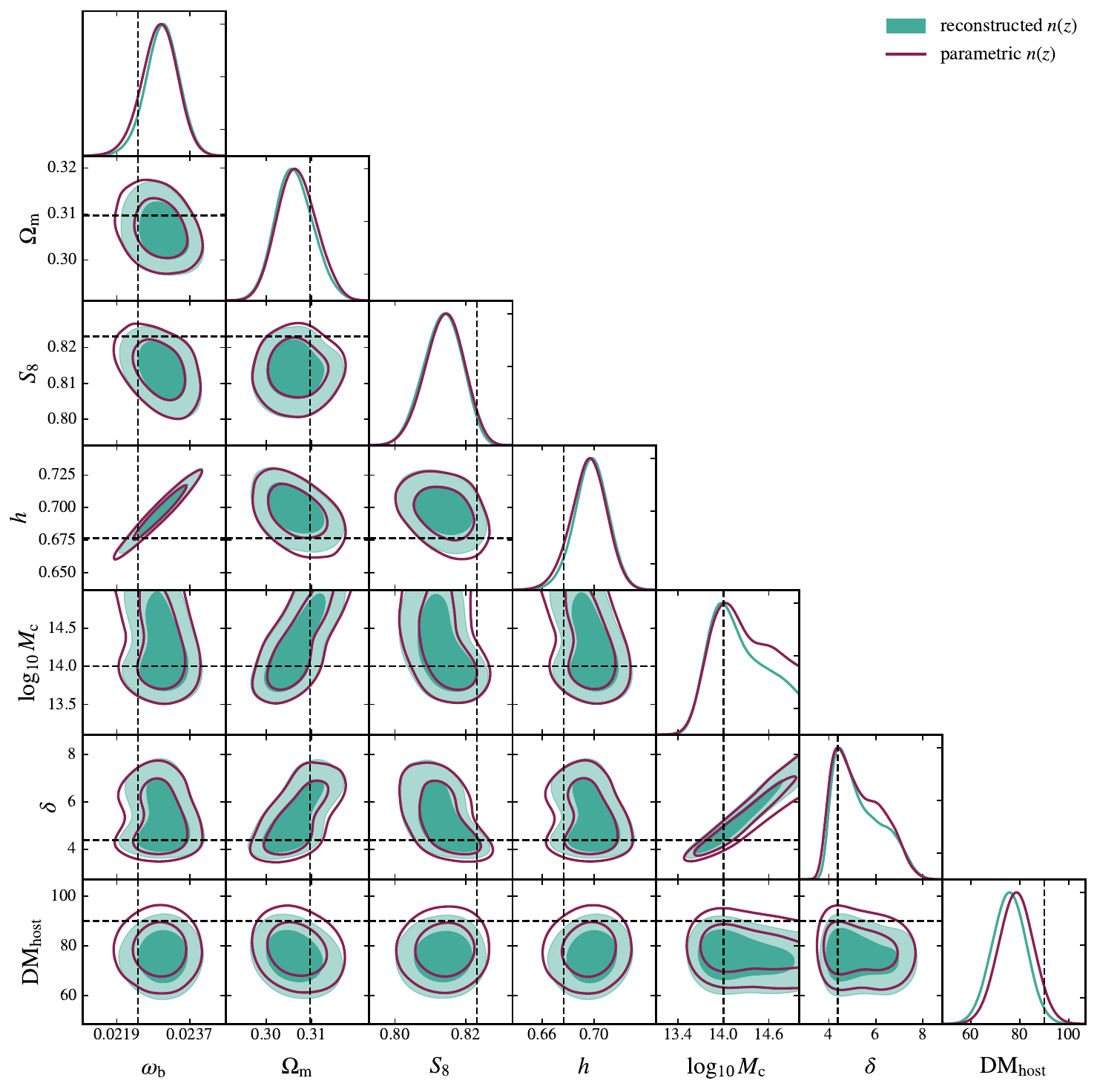}
      \caption{Recovery of the input parameters (dashed lines) from one noisy realisation of the forecast data vector. The filled contours show the case where $n(z)$ of the angular power spectra is reconstructed at each step by inverting the DMs of the host-identified FRBs via Equation~\eqref{eq:DM_i_obs}. The unfilled contours show the same data fitted with the parametric $n(z)$ of Section~\ref{ssec:forecast.survey}. Contours enclose the 68 and 95 per cent credible regions.}
      \label{fig:nzrecon}
  \end{figure}

  Since the uncertainty on $C_\ell^{\mathcal{D}\mathcal{D}}$ at large multipoles is largely set by the catalogue's shot noise, this bias grows with the number of FRBs. We therefore adopt an optimistic catalogue of $5 \times 10^5$ FRBs in Figure~\ref{fig:redshift_reconstruction}, comparable to the most ambitious samples expected from next-generation surveys. For a catalogue of the size considered in Section~\ref{ssec:forecast.survey}, we expect even the bias from a mismatched Macquart relation to lie well below the measurement uncertainty. For a fixed modelling bias and survey area, however, the uncertainty on $C_\ell^{\mathcal{DD}}$ scales approximately as $1/\bar{n}$ in the regime dominated by shot noise, so the significance of the bias grows linearly with sample size. A joint analysis is therefore not essential at the precision attainable with catalogues of the size considered in Section~\ref{ssec:forecast.survey}, but becomes increasingly important as FRB catalogues grow. This is precisely the regime in which the angular power spectrum is expected to become a competitive probe of baryonic feedback.

  In a joint analysis, the mean relation fitted to the host-identified subsample and the redshift distribution used to model the angular power spectra are described by the same parameters, here in particular $\mathrm{DM}_\mathrm{host}$, so the host-identified FRBs calibrate the redshift distribution of the full catalogue. We test this self-calibration directly on a single noisy realisation of the forecast data vector. We draw the bandpowers and the DMs of the 1444 host-identified FRBs from the fiducial model, keep the individual DMs, and bin them as before to form the Macquart block. We then fit these data as before, except that the parametric $n(z)$ of the angular power spectra is replaced by a reconstruction: at every step of the chain, we invert the unbinned DMs using the mean relation $\bar{\mathcal{D}}(z)$ and $\mathrm{DM}_{\rm host}$ of that step, and take the histogram of the resulting $\hat{z}_i$ as $n(z)$. In this test, the host-identified DMs serve as a proxy for the DMs of the FRBs without an identified host: their spectroscopic redshifts are not used in the reconstruction, and since they are drawn as a random subset of the catalogue, they sample the same redshift distribution. 

  Figure~\ref{fig:nzrecon} compares the resulting posterior with that obtained from the same data using the parametric $n(z)$. Both recover the input values within the scatter expected for a single realisation, and they agree within $0.5\sigma$ on every parameter. The remaining offsets from the input values, most visibly in $\mathrm{DM}_{\rm host}$, appear in both fits and therefore originate from the noise realisation. Reconstructing $n(z)$ through the jointly fitted mean relation thus introduces no bias at this sample size, provided the mean relation itself is modelled correctly (Figure~\ref{fig:redshift_reconstruction}). In a real analysis, the inversion would be applied to the FRBs without an identified host, which increases the number of objects entering the histogram but leaves the procedure unchanged. If the host-identified subsample were fully representative, its spectroscopic redshifts would directly constrain $n(z)$.
  In practice, host identification depends on redshift and burst properties, so inverting the DMs of the full sample through a jointly fitted mean relation is the more general route.

\section{Conclusions} \label{sec:conclusions}
 A catalogue of localised fast radio bursts (FRBs) provides two complementary observables: the dispersion measures (DMs) of individual bursts as a function of redshift, known as the Macquart relation, and the angular power spectrum of the DM field. As both observables are constructed from the same realisation of the electron density field, their joint analysis requires the cross-covariance between them, in addition to their respective auto-covariances. In this work, we have derived this cross-covariance, identified its physical origin, and quantified its impact on parameter inference.

 The cross-covariance is a connected three-point function sourced by the bispectrum of the electron overdensity, whereas the two auto-covariances are determined by two-point statistics and, in the case of the power spectrum, the connected trispectrum. We have evaluated the cross-covariance within the halo model and the Limber approximation. Statistical isotropy implies that only the monopole of the DM field correlates with its angular power spectrum, making the cross-covariance analogous to super-sample covariance: the DM of an individual burst probes the background mode, to which the power spectrum also responds. This interpretation leads to an exact bound on the correlation coefficient, which holds to all orders in the field and is independent of the bispectrum amplitude. The bound therefore provides a model-independent explanation for the low amplitude of the cross-covariance. In particular, since the variance of an individual DM is dominated by high multipoles, the monopole contributes only a small fraction of the total variance, consequently suppressing the correlation.

 We have validated our calculation using $5000$ forward-simulated catalogues, each containing $10^4$ FRBs, generated from log-normal realisations of the electron field. Measuring both observables from the same realisations retains the cross-covariance and enables a direct comparison between the theory and the simulation. The predicted and measured cross-covariances agree: when averaged over the host-identified FRBs, the correlation coefficient satisfies $|r|<0.005$ in every band, while individual DM--bandpower entries are typically at the level of one per cent, with the largest reaching a few per cent. However, a small cross-covariance relative to the diagonal elements does not, by itself, establish that it can be neglected in parameter inference. We therefore tested its impact directly by drawing $10^5$ mock data vectors from the full covariance and analysing each with and without the cross-covariance. Neglecting this term changes the width of the $\chi^2$ distribution by $0.2$ per cent and changes the acceptance decision at $95$ per cent confidence for $0.5$ per cent of the realisations. Compressing the Macquart data into redshift-binned means, as in our forecast, raises the correlation coefficients to $|r| \approx 0.2$ for 1444 host-identified FRBs. Nevertheless, neglecting the cross-covariance changes the width of the $\chi^2$ distribution by only about one per cent. These results indicate that, for a survey of the size and footprint considered here, the Macquart relation and the DM angular power spectrum can be treated as effectively independent probes, allowing their likelihoods to be combined without explicitly modelling their cross-covariance.

 Using the validated covariance, we performed a joint forecast combining an LSST-like weak lensing survey with a mock FRB catalogue. The mean DM--redshift relation provides additional information in a limited subset of parameter directions. It tightens the constraint on the mean host contribution by a factor of $4.1$, since this is the only observable in the data vector sensitive to a redshift-dependent additive offset. It also improves the constraint on the physical baryon density by $32$ per cent, primarily along the direction of constant $\omega_{\rm b}/h$, to which the amplitude of the mean relation is sensitive, while leaving $\Omega_{\rm m}$, $S_8$, and $h$ essentially unchanged. The baryonification parameters are constrained approximately $5$ per cent more tightly, but this improvement is indirect. Since the modelled mean electron column density is set by $\Omega_{\rm b}$ alone, the modelled mean DM is independent of the feedback parameters (modulo the neutral gas fraction) by construction. Instead, the improved constraints arise because tightening the constraint on the mean host contribution reduces nuisance freedom in the angular power spectrum, where the host contribution is partially degenerate with the feedback parameters. The size of this gain is further limited by a strong degeneracy between $\log_{10} M_{\rm c}$ and $\delta$, with $\log_{10} M_{\rm c}$ bounded only from below within its prior.

 Even where the Macquart relation adds little information directly, a joint analysis remains essential for interpreting the DM angular power spectrum. The redshift distribution of FRBs without identified hosts must be reconstructed by inverting the mean DM--redshift relation. A 20 per cent error in this relation biases $C_\ell^{\mathcal{DD}}$ by up to $2\sigma$ per band for a catalogue of $5 \times 10^5$ FRBs. Fitting both observables with the same parameters lets host-identified FRBs calibrate the redshift distribution of the full catalogue, which becomes increasingly important as catalogues grow.

 The following limitations are particularly relevant here. First, although the log-normal shells used for validation reproduce the one- and two-point statistics of the electron field by construction, their bispectrum does not exactly reproduce that of the halo model. The simulations therefore validate the predicted amplitude of the cross-covariance only to the extent that the log-normal bispectrum approximates the halo-model bispectrum. In this respect, the exact bound on the correlation coefficient, which is independent of the assumed bispectrum, provides a more general result. Second, the cross-covariance increases as the survey footprint decreases via $|r| \propto f_{\rm sky}^{-1/2}$. Extrapolating our measurements, it becomes relevant only for survey footprints of $\lesssim 150\,{\rm deg}^2$, which is far smaller than those of upcoming surveys.

 Furthermore, our forward simulations are idealised and do not yet account for realistic survey effects. A natural extension would incorporate the instrument's angular selection function, models of Milky Way and host-galaxy DM contributions, the preferential occurrence of FRBs in overdense regions, and the reduced detectability of FRBs with high DMs (see \cite{2020PhRvD.102b3528R} for some preliminary work). These effects can bias the measured auto-correlations and modify their covariance. Robust predictions for cross-correlations with other tracers would additionally require simulations that capture the non-linear relationship between the electron and matter fields, such as baryonified $N$-body lightcones.

 The effective independence of the mean DM--redshift relation and the DM angular power spectrum is a useful result for next-generation FRB surveys. For sufficiently wide footprints and at the current sample sizes, the two observables can be modelled and analysed separately, with their information combined at the likelihood level. Nevertheless, the cross-covariance derived here may become relevant as survey footprints shrink, DM measurements become more precise, or the number of host-identified FRBs per redshift bin grows. Its physical interpretation, model-independent bound, and explicit expression provide a basis for assessing its importance in future analyses.

\acknowledgments
 AW is supported by a Science and Technology Facilities Council studentship. We made extensive use of computational resources at the University of Oxford Department of Physics, funded by the John Fell Oxford University Press Research Fund.
 
\appendix
 \section{Spherical harmonic basis}\label{app:spherical_harmonic_basics}
  Plane waves can be expanded into spherical coordinates via:
  \begin{equation} \label{eq:rayleigh_expansion}
      e^{\mathrm{i}\mathbf{k}\cdot(\chi\hat{\mathbf{n}})} = 4\pi \sum_{\ell m} \mathrm{i}^\ell j_\ell(k\chi) \, Y_{\ell m}(\hat{\mathbf{n}}) \, Y_{\ell m}^*(\hat{\mathbf{k}}),
  \end{equation}
  which allows the expression of $\delta_{\rm e}(\chi(z)\hat{\mathbf{n}},z)$ as
  \begin{equation}
      \delta_{\rm e}(\chi(z)\hat{\mathbf{n}},z) = 4\pi \int \frac{\mathrm{d}^3 k}{(2\pi)^3} \, \delta_{\rm e}(\mathbf{k},z) \sum_{\ell m} \mathrm{i}^\ell j_\ell(k\chi) \, Y_{\ell m}(\hat{\mathbf{n}}) \, Y_{\ell m}^*(\hat{\mathbf{k}}).
  \end{equation}
  from which the spherical harmonic coefficients can be read off:
  \begin{equation}
      a_{\ell m}^{\mathcal{D}}(z) = 4\pi \mathrm{i}^\ell \int_0^{z} \mathrm{d}z' \, W_\mathcal{D}(z') \int \frac{\mathrm{d}^3 k}{(2\pi)^3} \, j_\ell(k\chi(z')) \, Y_{\ell m}^*(\hat{\mathbf{k}}) \, \delta_{\rm e}(\mathbf{k},z').
  \end{equation}
  It follows some useful relations for spherical harmonics. The orthogonality relation:
  \begin{equation}
      \int \mathrm{d}\hat{\mathbf{n}} \, Y_{\ell m}^*(\hat{\mathbf{n}}) \, Y_{\ell' m'}(\hat{\mathbf{n}}) = \delta_{\ell\ell'}^\mathrm{K} \, \delta_{mm'}^\mathrm{K},
  \end{equation}
  and
  \begin{equation}
  \label{app:eq:legendre}
      \sum_m Y_{\ell m}(\hat{\mathbf{k}}_3) \, Y_{\ell m}^*(\hat{\mathbf{k}}_2) = \frac{2\ell+1}{4\pi} \, P_\ell(\hat{\mathbf{k}}_2 \cdot \hat{\mathbf{k}}_3).
  \end{equation}

 \section{Density profiles for the halo model} \label{app:sec:profiles}
  Here, we describe the density profiles entering the halo model calculations of Section~\ref{sec:theory}. We adopt the baryonification (BFC) framework of Ref.~\cite{Schneider2025}, which extends the original method of Ref.~\cite{Schneider2015}. In BFC, the matter distribution of a gravity-only $N$-body simulation is modified by radially displacing particles so that they reproduce physically motivated density profiles for dark matter, gas, and stars. We split each halo into a dark matter component and a baryonic component, and further decompose the baryons into hot gas (hga), cold inner gas (iga), central galaxies (cga), and satellite galaxies (sga). Throughout, halo masses and radii are defined with respect to 200 times the critical density.

  Initially, dark matter and baryons follow the same spatial profile, differing only by their cosmic mass fractions, $f_{\rm dm} = \Omega_{\rm c}/\Omega_{\rm m}$ and $f_{\rm bar} = \Omega_{\rm b}/\Omega_{\rm m}$:
  \begin{align}
      \rho_{{\rm dm},i} &= f_{\rm dm}[\rho_{\rm nfw} + \rho_{\rm 2h}], \;\; \rho_{{\rm dm},f} = \rho_{\rm dm} + f_{\rm dm} \rho_{\rm 2h}, \nonumber \\
      \rho_{{\rm bar},i} &= f_{\rm bar}[\rho_{\rm nfw} + \rho_{\rm 2h}], \;\; \rho_{{\rm bar},f} = \rho_{\rm gas} + \rho_{\rm star} + f_{\rm bar} \rho_{\rm 2h}.
  \end{align}
  Here, the initial one-halo profile $\rho_{\rm dm}$ is a truncated NFW profile \cite{Navarro1997, Baltz2007, Oguri2011}. The final baryonic density is decomposed into gas and stellar components,
  \begin{equation}
      \rho_{\rm gas} = \rho_{\rm hga} + \rho_{\rm iga}, \qquad \rho_{\rm star} = \rho_{\rm cga} + \rho_{\rm sga}.
  \end{equation}

  The hot gas density profile is given by 
  \begin{equation} \label{eq:rho_hga}
      \rho_{\rm hga}(r) = \rho_{\rm hga,0}\left[1+x_{\rm c}^\alpha\right]^{-\beta(M_{200})/\alpha} \left[1+x_{\rm t}^{\gamma}\right]^{-\delta/\gamma},
  \end{equation}
  where $x_{\rm c} = r/r_{\rm c}$, $r_{\rm c} = \theta_{\rm co} r_{200}$, $x_{\rm t} = r/r_{\rm t}$, and $r_{\rm t} = \varepsilon r_{200}$. The exponents $\alpha = 1$ and $\gamma = 3/2$ set the sharpness of the transition at the core and truncation radii, while $\delta$ controls the steepness of the outer truncation, and hence the extent to which the gas extends beyond $r_{\rm 200c}$. The mass-dependent intermediate slope is
  \begin{equation}
      \beta(M_{200}) = \frac{3(M_{200}/M_{\rm c})^\mu}{1+(M_{200}/M_{\rm c})^\mu}.
  \end{equation}
  Haloes with $M_{\rm 200c} \ll M_{\rm c}$ have shallow gas profiles because feedback has expelled gas from their central regions, while haloes with $M_{\rm 200c} \gg M_{\rm c}$ retain steeper profiles with $\beta \to 3$. The parameter $\mu$ controls the sharpness of this transition. The normalisation $\rho_{\rm hga,0}$ is fixed such that the hot gas mass integrated to infinity equals $f_{\rm hga} M_{\rm tot}$, where
  \begin{equation} \label{eq:f_hga}
      f_{\rm hga}(M_{\rm 200c}) = \frac{\Omega_{\rm b}}{\Omega_{\rm m}} - f_{\rm star}(M_{\rm 200c}) - f_{\rm iga}(M_{\rm 200c}).
  \end{equation}
  Here, $f_{\rm star}$ is the stellar fraction, parametrised following abundance matching~\cite{Moster2013, Schneider2019}, and $f_{\rm iga}$ is the cold gas fraction, taken to be proportional to the central galaxy fraction~\cite{Schneider2025}. Feedback therefore redistributes gas within and beyond the halo, but does not change its total gas content.

  Assuming the gas is fully ionised, the electron number density is
  \begin{equation} \label{eq:ne}
      n_{\rm e}(r\,|\,M) = \frac{1+x_{\rm H}}{2m_{\rm p}}\, \rho_{\rm gas}(r\,|\,M),
  \end{equation}
  where $x_{\rm H} = 0.76$ is the hydrogen mass fraction. Since the prefactor is constant, it cancels in the electron overdensity, and the profile entering the halo model is
  \begin{equation} \label{eq:ue}
      u_{\rm e}(k\,|\,M) = \frac{4\pi}{\bar{\rho}_{\rm gas}} \int_0^\infty\mathrm{d}r\, r^2\, j_0(kr)\, \rho_{\rm gas}(r\,|\,M),
  \end{equation}
  where $\bar{\rho}_{\rm gas}$ is the mean gas density computed from the halo model.

  We fix $\theta_{\rm co} = 0.3$, $N_{\rm star} = 0.03$, $\eta = 0.1$, ${\rm d}\eta = 0.22$ to the values of Ref.~\cite{Kovac2025}, and adopt the fiducial values $\log_{10}(M_{\rm c}/M_\odot) = 14.0$, $\mu = 0.74$, and $\delta = 4.38$.

 \section{Halo model description of the electron density field} \label{app:sec:halo_model}
  We model the electron power spectrum, bispectrum, and trispectrum required for the covariance calculation within the halo model. We write the Fourier space profile of the electron field as $u_{\rm e}(k,z|M)$ for a halo of mass $M$ at redshift $z$. We define the halo model integrals
  \begin{equation}
      I_\nu^\beta(k_1, \dots, k_\nu, z) = \int\mathrm{d}M\, n(M,z)\, b_\beta(M,z)\, \left\langle \prod_{i=1}^\nu u_{\rm e}(k_i,z|M) \right\rangle,
  \end{equation}
  where $n(M,z)$ is the halo mass function, $b_\beta(M,z)$ is the halo bias of order $\beta$, with $b_0=1$ and $b_1=b$, and the brackets denote the average over the electron profile at fixed halo mass. For the profiles considered here, we approximate this average as a product,
  \begin{equation}
      \left\langle \prod_{i=1}^\nu u_{\rm e}(k_i,z|M) \right\rangle = \prod_{i=1}^\nu u_{\rm e}(k_i,z|M).
  \end{equation}
  In particular,
  \begin{equation}
      I_1^0 = \int\mathrm{d}M\, n(M,z)\, u_{\rm e}(k,z|M),
  \end{equation}
  \begin{equation}
      I_1^1 = \int\mathrm{d}M\, n(M,z)\, b(M,z)\, u_{\rm e}(k,z|M),
  \end{equation}
  with analogous expressions for the higher-order integrals.

  The electron power spectrum entering the DM two-point functions is written as the sum of one- and two-halo contributions,
  \begin{equation}
      P_{\rm e}(k,z) = P_{\rm e}^{\rm 1h}(k,z) + P_{\rm e}^{\rm 2h}(k,z),
  \end{equation}
  where
  \begin{equation}
      P_{\rm e}^{\rm 1h}(k,z) = I_2^0(k,k,z),
  \end{equation}
  \begin{equation}
      P_{\rm e}^{\rm 2h}(k,z) = \left[I_1^1(k,z)\right]^2\, P_{\rm lin}(k,z).
  \end{equation}
  Similarly, the electron bispectrum required for the DM--$C_\ell^{\mathcal{DD}}$ cross-covariance is decomposed into one-, two-, and three-halo contributions,
  \begin{equation}
      B_{\rm eee} =  B^{\rm 1h}_{\rm eee} + B^{\rm 2h}_{\rm eee} + B^{\rm 3h}_{\rm eee},
  \end{equation}
  where
  \begin{equation}
      B^{\rm 1h}_\mathrm{eee}(k_1,k_2,k_3,z) = I_3^0(k_1,k_2,k_3,z),
  \end{equation}
  \begin{equation}
      B^{\rm 2h}_{\rm eee}(k_1,k_2,k_3,z) = I_1^1(k_1) I_2^1(k_2,k_3) P^{\rm lin}(k_1) + \;\mathrm{perm.},
  \end{equation}
  \begin{equation}
      B^{\rm 3h}_{\rm eee}(k_1,k_2,k_3,z) = \prod_{i=1}^3 I^1_1(k_i,z) B_{\rm tree}(k_1,k_2,k_3,z).
  \end{equation}
  Here,
  \begin{equation}
      B_{\rm tree}(\mathbf{k}_1,\mathbf{k}_2,\mathbf{k}_{3}) = 2F^{\rm s}_2(\mathbf{k}_1,\mathbf{k}_2) P_{\rm lin}(k_1) P_{\rm lin}(k_2) + 2\;\mathrm{perm.}
  \end{equation}
  is the tree-level matter bispectrum, which in turn depends on the symmetrised second-order perturbation theory kernel $F^{\rm s}_2$. This bispectrum provides the leading non-Gaussian contribution to the cross-covariance between an individual FRB and the angular power spectrum.

  The connected four-point function of the electron overdensity field determines the non-Gaussian contribution to the angular power spectrum covariance. Within the halo model, this is decomposed as
  \begin{equation}
      T_{\rm eeee} = T^{\rm 1h}_{\rm eeee} + T^{\rm 2h}_{\rm eeee} + T^{\rm 3h}_{\rm eeee} + T^{\rm 4h}_{\rm eeee}.
  \end{equation}
  The one-halo contribution is
  \begin{equation}
      T_{\rm eeee}^{\rm 1h}(k_1, k_2, k_3, k_4, z) = I_4^0(k_1, k_2, k_3, k_4, z).
  \end{equation}
  We construct the two-, three-, and four-halo contributions from the same halo model integrals, along with the appropriate halo bias factors and perturbative matter polyspectra. We use the resulting trispectrum to calculate the isotropised connected trispectrum $\bar{T}_{\rm eeee}$ that enters the non-Gaussian angular power spectrum covariance. 
  
  The halo model therefore provides a consistent hierarchy of correlation functions for the covariance calculation. The electron power spectrum determines the DM two-point statistics and the Gaussian covariance, the electron bispectrum determines the cross-covariance between an individual DM and the angular power spectrum, and the connected electron trispectrum determines the non-Gaussian contribution to the $C_\ell^{\mathcal{DD}}$ covariance. In this way, a single halo model framework describes all entries of the joint covariance matrix.

\bibliographystyle{JHEP}
\bibliography{references}

\end{document}